\documentstyle[amssymb,11pt]{article}
\input epsf
\input{tcilatex}
\begin{document}

\hfill CPTH S.636 0798

\hfill hepth/9808004

\hfill July, 1998

\vspace{20pt}

\begin{center}
{\large {\bf \ }}

{\large {\bf THEORY\ OF\ HIGHER SPIN TENSOR\ CURRENTS }}

{\large {\bf AND CENTRAL\ CHARGES}}
\end{center}

\vspace{6pt}

\begin{center}
{\sl Damiano Anselmi}

{\it Centre de Physique Theorique, Ecole Polytechnique, F-91128 Palaiseau
Cedex, FRANCE}~{\sl \footnote{{\sl {\rm {Address starting Oct. 1$^{st}$
1998: Theory Group, CERN, Geneva, Switzerland.}}}}}
\end{center}

\vspace{8pt}

\begin{center}
{\bf Abstract}
\end{center}

We study higher spin tensor currents in quantum field theory. Scalar, spinor
and vector fields admit unique ``improved'' currents of arbitrary spin,
traceless and conserved. Off-criticality as well as at interacting fixed
points conservation is violated and the dimension of the current is
anomalous. In particular, currents ${\cal J}^{(s,I)}$ with spin $0\leq s\leq
5$ (and a second label $I$) appear in the operator product expansion of the
stress tensor. The $TT$ OPE\ is worked out in detail for free fields;
projectors and invariants encoding the space-time structure are classified.
The result is used to write and discuss the most general OPE for interacting
conformal field theories and off-criticality. Higher spin central charges $%
c_{s}^{I}$ with arbitrary $s$ are defined by higher spin channels of the
many-point $T$ -correlators and central functions interpolating between the
UV\ and IR limits are constructed. We compute the one-loop values of all $%
c_{s}^{I}$ and investigate the RG trajectories of quantum field theories in
the conformal window following our approach. In particular, we discuss
certain phenomena (perturbative and nonperturbative) that appear to be of
interest, like the dynamical removal of the $I$-degeneracy. Finally, we
address the problem of formulating an action principle for the RG\
trajectory connecting pairs of CFT$_{4}$'s as a way to go beyond
perturbation theory.

\vfill\eject

\section{Introduction and motivation.}

In quantum field theory, particularly important for a theoretical
investigation of phenomena beyond perturbation theory, but not so far from
it, is the so-called ``conformal window''. In QCD\ with $N_{c}$ colors and $%
N_{f}$ flavors the theory has a weakly coupled IR fixed point in the
neighborhood $N_{f}\lesssim 11/2~N_{c}$, where the large distance limit of
the theory can be studied to the first orders in the perturbative expansion.
By continuity, there is a certain finite interval sharing similar properties
as this infinitesimal neighborhood. Its lower bound $N_{f~\min }$ is not
known. $N_{f~\min }<N_{f}<11/2~N_{c}$ is the conformal window, where the
theory has an IR\ fixed point and, by definition, one can reach the IR fixed
point by resumming the perturbative expansion. Below $N_{f~\min }$ purely
non-perturbative effects become important. QCD\ does not belong to the
conformal window, nevertheless a better understanding of the conformal
window is a natural first step to go beyond perturbation theory. In
supersymmetric theories, on the other hand, one knows the precise size of
the conformal window in various cases and there exist duality arguments in
favor of the equivalence of the IR\ limits of very different theories \cite
{seiberg}. Moreover, rigorous computations of IR quantities have been
performed \cite{noi,noi2}.

In the conformal window, the renormalization group flow can be thought of as
the radiative interpolation between two conformal field theories (CFT), the
UV\ and IR\ limits of the renormalization group (RG) flow\ trajectory. The
purpose of this paper is to make a further step in the program addressed in
ref. \cite{ccfis}, which indeed amounts to study quantum field theory (QFT)\
under this point of view. We refine the research purpose itself. The reader
interested in the chronological development of this program up to now should
consult, in the order, references \cite{noialtri,c',ccfis,noi,noi2}.

In the present paper, we develop techniques similar to those used in
the context of deep inelastic scattering \cite{muta} and apply them to 
study the operator product expansion of the stress-energy tensor
in the Euclidean framework.
We thus clarify the nature of the ``secondary'' central charges introduced
in \cite{noialtri,c',ccfis} and study several phenomena in connection with
this issue. The stress tensor OPE
is worked out in detail for free fields and generalized to
interacting conformal field theories and off-criticality. This analysis
leads us to consider higher spin currents, which indeed appear as channels
of the $TT$ OPE.

We investigate various properties of higher spin tensor currents in the
context of our approach to quantum field theory. We can define
and study the OPE's of these currents
and their anomalies, the RG\ flow of certain ``central
functions'', defined by the two- and four-point functions of the currents in
exam and by their trace anomalies, and compute their UV\ and IR critical\
limits (central charges). For spin 1 and 2 this program was successfully
realized in \cite{noi} for supersymmetric theories in the conformal window.
In particular, the exact IR\ values of the gravitational central charges
(called $c$ and $a$) were computed for UV\ free theories using the
remarkable properties of supersymmetry. The strategy was applied in \cite
{noi2} to a large amount of models. The results of \cite{noi,noi2} provided,
among the other things, a strong support to the idea of irreversibility of
the flux of the RG\ flow (``$a$-theorem'').

The question that we address now is: what can we say in the same spirit
using higher spin currents? First of all, the assumption that a quantum
field theory admits a higher spin flavor symmetry is not met by ordinary
theories: interactions violate explicitly the higher spin conservation law
and higher spin currents acquire an anomalous dimension already at the first
orders in the perturbative expansion. For this reason it is not sufficient
to study two-point functions of higher spin currents, rather one has to look
for higher spin channels inside the correlators of the stress tensor \cite
{ccfis}. To begin with, one has to study the $TT$ OPE in detail.

``Improved'' higher spin tensor currents provide a basis
for the terms appearing in the $TT$ OPE (this and other statements will be
made precise in the paper) and provide a simple way to classify
the primary operators in the OPE. The construction of certain projectors
that encode the space-time structure of the terms allows us to generalize
the OPE to interacting conformal field theories and off-criticality.
Equipped with this, one can investigate the OPE along the RG trajectory of a
quantum field theory.

Along the RG\ flow trajectory, higher spin currents ``move'',\ due to their
anomalous dimension. The phenomenology of this moving is an interesting
subject to study in this domain. In particular, this phenomenon removes a
certain degeneracy that one observes in the free field limit. Examples will
be considered in detail, in N=4 supersymmetric Yang-Mills theory as well as
in the context of electric-magnetic duality \cite{seiberg}.

Before starting the technical analysis, we would like to formulate the idea
underlying our approach to QFT in general terms.

One can naturally view perturbation theory as a ``Cauchy problem''. The
starting conformal field theory, say CFT$_{UV}$ (that we assume free), and
the vertices of the classical Lagrangian are, so to speak, the {\sl initial
conditions } (``position'' and ``velocity'', respectively). The functional
integral is the step-by-step algorithm to move towards an unknown CFT$_{IR}$%
. The problem is then to identify CFT$_{IR}$ given the initial conditions.

\vskip .5truecm

%%Begin InstantTeX Picture
\let\picnaturalsize=N

%If you do not have the picture file add:
%\let\nopictures=Y
%to the beginning of the file.
\ifx\nopictures Y\else{\ifx\epsfloaded Y\else\fi
\global\let\epsfloaded=Y \centerline{\ifx\picnaturalsize N\epsfxsize
4.0in\fi \epsfbox{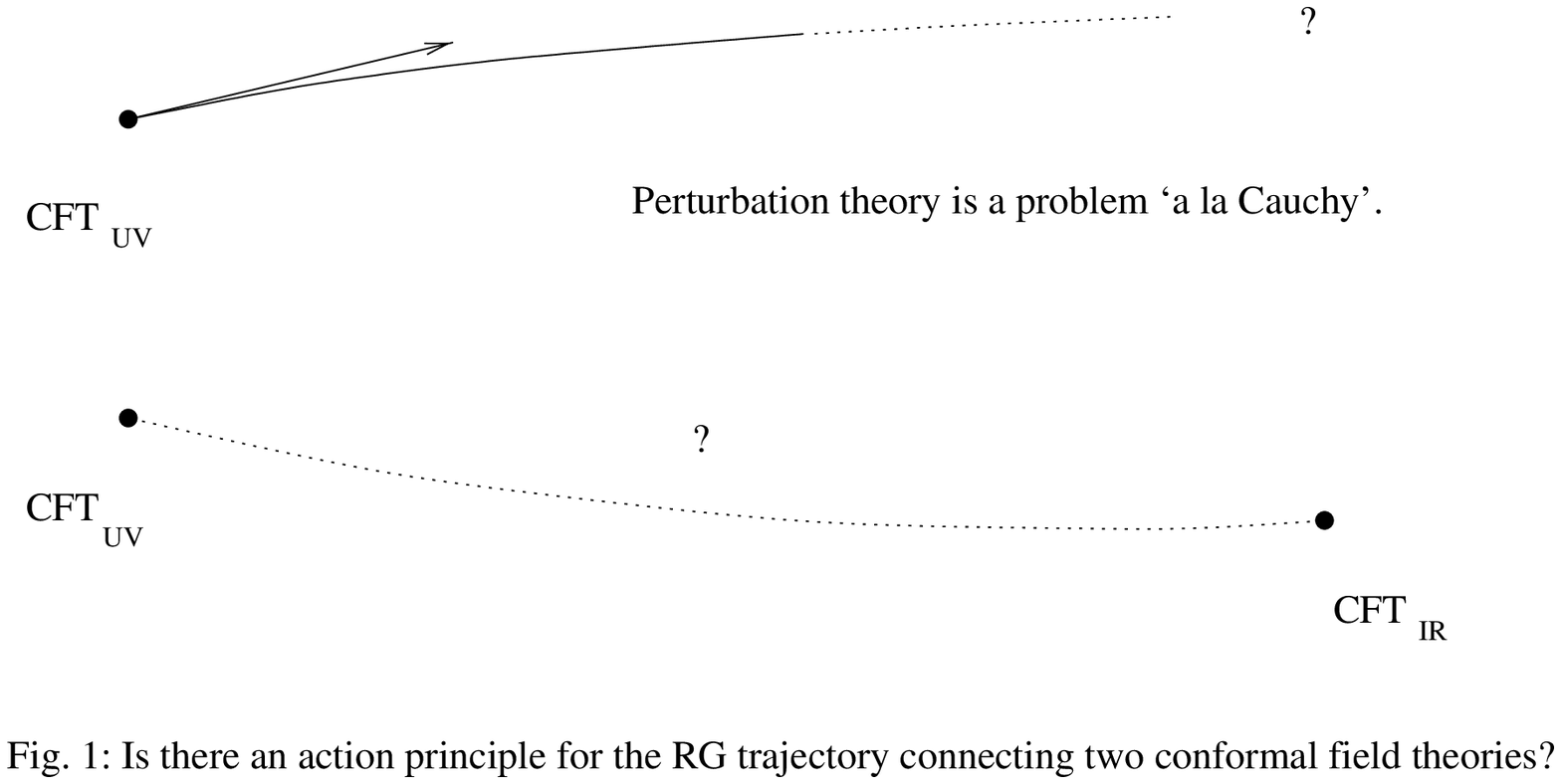}}}\fi
%%End InstantTeX Picture
\vskip .3truecm

Alternatively, a trajectory can be identified by its initial and final
limits and by the requirement that a certain functional (``action'') have
its minimum value with these boundary conditions. One is therefore led to
ask:\ is there an action principle identifying {\sl the} quantum field
theory (i.e. the RG\ trajectory), hopefully unique under suitable
assumptions, that radiatively interpolates between two {\sl given} CFT's?\
Presumably one will be led to consider a set of theories larger than the set
of ordinary renormalizable theories.

We are not ready, yet, to answer this difficult question. The best that we
can do is to reconsider ordinary theories, as we know them from perturbation
theory, under this alternative point of view (see Fig. 1) and see what we
can learn. The properties that we derive might become axioms of the new
formulation or maybe stimulate a numerical research.

First we have to fix the set of quantities that should be studied in this
approach. A quantum field theory is identified by the set of its
correlators, both of elementary fields and composite operators. However, in
general there is a undesirable dependence on the description (what are the
elementary fields, quarks and gluons or baryons and mesons? and what is
composite?), the definition (a composite operator, as well as an elementary
field, needs to be normalized at some energy scale, otherwise one can
multiply it by an arbitrary function of the running coupling constant), and
so on. We look for quantities that are invariant under these details.
Introducing an intermediate reference scale does not change the RG
trajectory and so it is immaterial to our problem. Moreover, in the
reformulation of QFT\ as we imagine it, it should not be necessary to speak
about ``elementary'' fields, nor ``composite'' operators, since what is
elementary in the UV\ is not elementary in the IR\ and vice versa.

It is natural to focus on conserved currents, which do not need any
independent normalization, and study their correlators. Conserved currents
should be the ``elementary'' objects of our description, whatever fields are
used to represent them explicitly at a given energy scale. Among these, the
stress-tensor plays a peculiar role, since it always exists. The idea is to
define certain key-quantities (central charges) via correlators of conserved
currents, characterizing the theory in the conformal fixed point and to
interpolate between the UV\ and IR\ values of these quantities by suitable
functions. These functions should allow one to study the evolution of the
theory along the RG\ trajectory. Once these functions will be understood it
will be possible to make some proposal for the envisaged RG\ action
principle. 

We will study, among the other things, the $TT$ OPE and certain structures
of the multi-$T$-point-functions, which encode the central charges just
mentioned.

Some interesting phenomena occurring in this approach will be described in
the paper. Just to mention one example on which we will not come back again,
the irreversibility properties exhibited by some central functions (the ``$a$%
-functions'') would imply that, given the two extremal conformal field
theories of the RG trajectory, one knows {\sl a priori}, by comparing the
values of the $a$-functions in the two cases, which extremum is the UV
limit\ and which extremum is the IR\ limit. The $a$-functions should
determine a level (``potential'') that classifies conformal field theories
and puts restrictions on the possible RG\ trajectories connecting them. For
example, two N=4 supersymmetric Yang-Mills theories with the same gauge
group and different values of the coupling constant $g$ very presumably will
admit no interpolating RG\ trajectory. The explanation might be: because
they have the same $a$-values \cite{noi} and theories with the same $a$%
-values are only marginally connectable. Note that the converse has already
been proved \cite{c',noi}: marginal deformations do not change the $a$%
-values.

The paper is divided into four sections. In the first section (section 2)
the general form of improved higher spin currents and their critical
two-point functions are studied. In section 3 currents and two-point
correlators are extended off-criticality, first under the assumption of
higher spin flavor symmetry and then in presence of an anomalous dimension
(explicit violation of the conservation condition). In section 4 the $TT$
OPE is worked out in detail. In section 5 the results of the other sections
are combined in order to construct the desired central charges and central
functions. Certain phenomena are discussed in supersymmetric theories and
some simple remarks in the context of the AdS/CFT correspondence \cite{malda}
are made. Relationships with more familiar anomalies are pointed out.

\smallskip For notational convenience, let us write down, before beginning
the discussion of higher spin tensors currents, some simple spin-1 and
spin-2 formulas that we will generalize. The spin-1 current and two-point
function read in the free field limit 
\begin{equation}
j_{\mu }=\bar{\psi}\gamma _{\mu }\psi ,\qquad <j_{\mu }(x)~j_{\nu }(0)>={\rm %
const}.~\pi _{\mu \nu }\left( \frac{1}{|x|^{4}}\right) ,  \label{11}
\end{equation}
where $\pi _{\mu \nu }=\partial _{\mu }\partial _{\nu }-\Box \delta _{\mu
\nu }$ (the precise coefficients will be written in the relevant sections).
There is also the axial current 
\begin{equation}
{\cal A}_{\mu }^{(1)}=\bar{\psi}\gamma _{5}\gamma _{\mu }\psi =-\frac{1}{6}%
\varepsilon _{\mu \nu \rho \sigma }\overline{\psi }\gamma _{\nu \rho \sigma
}\psi .  \label{barba2}
\end{equation}
where $\gamma _{5}=\gamma _{1}\gamma _{2}\gamma _{3}\gamma _{4}.$ For spin-2
we have 
\begin{equation}
T_{\mu \nu }=\frac{2}{3}\partial _{\mu }\varphi \partial _{\nu }\varphi -%
\frac{1}{6}\delta _{\mu \nu }\left( \partial _{\alpha }\varphi \right) ^{2}-%
\frac{1}{3}\varphi \partial _{\mu }\partial _{\nu }\varphi ,\quad <T_{\mu
\nu }(x)~T_{\rho \sigma }(0)>={\rm const}.~{\prod }_{\mu \nu ,\rho \sigma
}^{(2)}\left( \frac{1}{|x|^{4}}\right) ,  \label{12}
\end{equation}
where 
\[
{\prod }_{\mu \nu ,\rho \sigma }^{(2)}={\frac{1}{2}}(\pi _{\mu \rho }\pi
_{\nu \sigma }+\pi _{\mu \sigma }\pi _{\rho \nu })-{\frac{1}{3}}\pi _{\mu
\nu }\pi _{\rho \sigma }. 
\]
This stress tensor is known as the ``improved'' stress tensor. For this
notion and references about it we point out the well-known lectures by
Jackiw \cite{jackiw}.

A discussion about higher spin currents can be found in ref. \cite{berends}.
We will use those results and other results of previous work on this
subject. The list of references contained here about higher spin fields
nevertheless incomplete. We are more concerned with the properties of higher
spin currents for lower spin fields (scalar, spinor and vector fields),
rather than the dynamical aspects of higher spin fields in themselves, to
which a great amount of effort has been nevertheless devoted by physicists
so far. In particular, recent developments towards a consistent formulation
of dynamical higher spin couplings have shown that the problem of coupling
higher spin fields to gravity can be overcome in presence of a cosmological
term and that an all-order consistent formulation of higher spin couplings
can be achieved at least at the level of equations of motion. For this and
related issues, we point out the recent paper by Vasiliev \cite{vasiliev},
which is also a review of the matter and contains a detailed list of
references and an historical survey.

\section{Higher spin tensor currents and their two-point functions.\label%
{ulk}}

A higher spin current ${\cal J}_{\mu _{1}\cdots \mu _{s}}^{(s)}(x)$ is a
completely symmetric operator in its $s$ indices, satisfying the double
trace condition \cite{fronsdal,dewit} 
\begin{equation}
{{{{{{\cal J}^{(s)}}_{\mu }}^{\mu }}_{\nu }}^{\nu }}_{\mu _{5}\cdots \mu
_{s}}=0  \label{doubletrace}
\end{equation}
and the traceless conservation condition 
\begin{equation}
\partial ^{\mu }{\cal J}_{\mu \mu _{2}\cdots \mu _{s}}^{(s)}-{\frac{1}{%
2(s-1)}}\sum_{i<j=2}^{s}\delta _{\mu _{i}\mu _{j}}\partial ^{\mu }{%
{\cal J}^{(s)}}_{\mu \cdots \mu _{i-1}\alpha \mu _{i+1}\cdots \mu
_{j-1}\alpha \mu _{j+1}\cdots \mu _{s}}=0.  \label{conservation}
\end{equation}
These properties make the interaction lagrangian 
\[
{\cal L}_{I}=h_{s}^{\mu _{1}\cdots \mu _{s}}{\cal J}_{\mu _{1}\cdots \mu
_{s}}^{(s)} 
\]
invariant under the gauge transformation 
\begin{equation}
\delta h_{s}^{\mu _{1}\cdots \mu _{s}}=\partial _{\{\mu _{1}}{\xi _{s}}%
_{\,\mu _{2}\cdots \mu _{s}\}},~~~~~~{{{\xi _{s}}_{\,\mu }}^{\mu }}_{\mu
_{3}\cdots \mu _{s}}=0,  \label{gaugeinv}
\end{equation}
with ${\xi _{s}}_{\,\mu _{2}\cdots \mu _{s}}$ symmetric in its $s-1$
indices. Counting the components and fixing the gauge, one can show that
higher spin fields have two elicities. The counting proceeds as follows. A $%
k $-index symmetric traceless tensor has $(k+1)^{2}$ components and a doubly
traceless tensor has $(k+1)^{2}+(k-1)^{2}=2(k^{2}+1)$ components. (\ref
{conservation}) is an $\left( s-1\right) $-indexed such tensor, and
therefore contains $s^{2}$ components. Fixing the gauge means to subtract
twice as many components (``ghosts'' and ``antighosts''), therefore giving $%
2(s^{2}+1)-2s^{2}=2$ surviving elicities, the correct number for a massless
spin-$s$ propagating field. One can write a kinetic action for $h_{s}^{\mu
_{1}\cdots \mu _{s}}$ that reduces to the Klein-Gordon equation for each
elicity upon fixing the gauge \cite{fronsdal,dewit}.

Here we consider higher spin tensor currents for scalar, spinor and vector
fields. Let us start from free fields. The form of the operator ${\cal J}%
_{\mu _{1}\cdots \mu _{s}}^{(s)}$, as specified by conditions (\ref
{doubletrace}) and (\ref{conservation}) and by its minimal dimension $2+s$,
is actually not unique. It is unique once one imposes a further condition of 
{\it simple} tracelessness: 
\begin{equation}
{{{{\cal J}^{(s)}}_{\mu }}^{\mu }}_{\mu _{3}\cdots \mu _{s}}=0.
\label{simpletrace}
\end{equation}

This condition is the analogue of the tracelessness condition $T_{\mu \mu
}=0 $ for the stress-energy tensor $T_{\mu \nu }=${{{${\cal J}$}}}${{{_{\mu
\nu }^{(2)}}}}$. As a consequence, it is evident that it is an {\it %
additional} condition and that it does not characterize the higher spin
field in itself. It should be related to an extra symmetry of the theory. It
is tempting to think that the simple trace condition is the effect of the
conformal symmetry, i.e. that condition (\ref{simpletrace}) can be imposed
only in the critical points (high energy limit or large distance limit) of a
quantum field theory. Off-criticality condition (\ref{simpletrace}) should
not hold. With condition (\ref{simpletrace}) the counting is modified as
follows. (\ref{simpletrace}) leaves the tensor with $(s+1)^{2}$ components
and (\ref{conservation}) subtracts $s^{2}$ components. Thus one remains with 
$2s+1$ components, which is correct for a spin-$s$ operator.

Let us assume that ${\cal J}^{(s)}$ is conserved and doubly traceless. We
are going to show that if the correlators of ${\cal J}^{(s)}$ are conformal,
then ${\cal J}^{(s)}$ is simply traceless and has no anomalous dimension.
Vice versa, if ${\cal J}^{(s)}$ is simply traceless and has no anomalous
dimension, then its correlators are conformal. Some arguments of this
subsection are also treated in ref. \cite{**}, to which the reader 
is referred for comparison.

To begin with, let us study the two-point function of ${\cal J}^{(s)}$.
Conformal invariance assures that the current transforms as \cite{gatto,**} 
\begin{equation}
{\cal J}_{\mu _{1}\cdots \mu _{s}}^{(s)}(x^{\prime })~=|x|^{2s+4+2h}{\cal I}%
_{\mu _{1}\mu _{1}^{\prime }}(x)\cdots {\cal I}_{\mu _{s}\mu _{s}^{\prime
}}(x)~{\cal J}_{\mu _{1}^{\prime }\cdots \mu _{s}^{\prime }}^{(s)}(x)
\label{trasfo}
\end{equation}
under the coordinate inversion $x_{\mu }^{\prime }=x_{\mu }/x^{2}$. $h$ is
an eventual anomalous dimension of the operator ${\cal J}$. Therefore the
two-point correlator transforms as 
\begin{eqnarray}
<{\cal J}_{\mu _{1}\cdots \mu _{s}}^{(s)}(x^{\prime })~{\cal J}_{\nu
_{1}\cdots \nu _{s}}^{(s)}(y^{\prime })> &=&|x|^{2s+4+2h}|y|^{2s+4+2h}~{\cal %
I}_{\mu _{1}\mu _{1}^{\prime }}(x)\cdots {\cal I}_{\mu _{s}\mu _{s}^{\prime
}}(x)~{\cal I}_{\nu _{1}\nu _{1}^{\prime }}(y)\cdots {\cal I}_{\nu _{s}\nu
_{s}^{\prime }}(y)\times  \nonumber \\
&<&{\cal J}_{\mu _{1}^{\prime }\cdots \mu _{s}^{\prime }}^{(s)}(x)~{\cal J}%
_{\nu _{1}^{\prime }\cdots \nu _{s}^{\prime }}^{(s)}(y)>.  \label{con}
\end{eqnarray}
To write down the general solution to this condition the basic ingredient is
the symmetric orthogonal matrix ${\cal I}_{\mu \nu }(x)=\delta _{\mu \nu
}-2x_{\mu }x_{\nu }/x^{2},$ which is proportional to the Jacobian $\partial
x_{\mu }^{\prime }/\partial x_{\nu }$ and satisfies the properties

\begin{equation}
{\cal I}_{\mu \nu }(x^{\prime }-y^{\prime })={\cal I}_{\mu \mu ^{\prime }}(x)%
{\cal I}_{\mu ^{\prime }\nu ^{\prime }}(x-y){\cal I}_{\nu ^{\prime }\nu
}(y),\qquad {\cal I}_{\mu \nu }(x^{\prime })={\cal I}_{\mu \nu }(x),\qquad 
{\cal I}_{\mu \rho }(x){\cal I}_{\rho \nu }(x)=\delta _{\mu \nu }.
\label{orto1}
\end{equation}
Our correlator can then be written as\footnote{%
See for example \cite{c'} for a three-point correlator constructed with this
technique.} 
\begin{eqnarray}
<{\cal J}_{\mu _{1}\cdots \mu _{s}}^{(s)}(x)~{\cal J}_{\nu _{1}\cdots \nu
_{s}}^{(s)}(0)>=\frac{1}{|x|^{4+2s+2h}}\sum_{n=0}^{[s/2]} &a_{n}&\sum_{{\rm %
symm}}\delta _{\mu _{1}\mu _{2}}\delta _{\nu _{1}\nu _{2}}\cdots \delta
_{\mu _{n-1}\mu _{n}}\delta _{\nu _{n-1}\nu _{n}}\times  \nonumber \\
&&{\cal I}_{\mu _{n+1}\nu _{n+1}}(x)\cdots {\cal I}_{\mu _{s}\nu _{s}}(x).
\label{confo}
\end{eqnarray}
$[s/2]$ denotes the integral part. $\sum_{{\rm symm}}$ means complete
symmetrization in both sets of indices, i.e. the sum over the necessary
permutations divided by the number of permutations. $a_{n}$ are arbitrary
constants. It is straightforward to verify that the correlator transforms
correctly under inversion, since each term in the sum does. The basic rule
is that pairs of indices belonging to the same current, like $\mu _{1}\mu
_{2}$, are coupled via the identity matrix $\delta _{\mu _{1}\mu _{2}},$
while pairs of indices belonging to different currents, like $\mu _{1}\nu
_{1},$ are coupled via the matrix ${\cal I}$: ${\cal I}_{\mu _{1}\nu _{1}}$.
One then writes down all possible terms. The conclusion is that there are $%
[s/2]+1$ free parameters and one anomalous dimension.

We now study the imposition of (\ref{doubletrace}), (\ref{conservation}) and
(\ref{simpletrace}).

The double trace condition (\ref{doubletrace}) imposes $[s/2]-1$ conditions.
First, one notes that doubly tracing a conformal correlator (\ref{con}) one
still obtains a conformal correlator. Therefore one can repeat the previous
counting of allowed terms with the reduced set of indices.

After imposing the double trace condition, only two free parameters survive
(one of which is the overall constant), plus the eventual anomalous
dimension $h$. Finally, conservation (\ref{conservation}) kills another
parameter and furthermore imposes $h=0$. One remains just with the overall
constant. For the applications it is important to remark that the dimension
is fixed to be the canonical one. One can check that the resulting
expression automatically satisfies the simple trace condition (\ref
{simpletrace}).

Now we proceed differently. We start from the conformal correlator (\ref{con}%
) and impose the double trace condition, therefore remaining with two free
parameters and the anomalous dimension $h$. Then, instead of imposing the
conservation condition, we just impose the simple trace condition (\ref
{simpletrace}). The result is that one of the two parameters is killed, but $%
h$ remains free: we remain with the overall constant and the anomalous
dimension $h$: $h$ is not necessarily zero and when it is nonzero the
conservation condition (\ref{conservation}) is violated.

The simple trace condition can always be imposed on the current ${\cal J}%
_{\mu _{1}\cdots \mu _{s}}^{(s)},$ since it is a purely algebraic condition.
We will show that in general the conservation condition cannot be satisfied.
In particular, it cannot be satisfied in interacting CFT$_{4}$'s. We
conclude that $h$ measures the violation of the conservation condition at
criticality.

Anomalous violations of the conservation conditions can usually be moved
away with the so-called finite local counterterms. For example, the
violation of the vector current conservation can be moved to the divergence
of the axial current \cite{jackiw}, the violation of the stress current
conservation can be moved to the trace condition.

The $h$-violation of the higher spin conservation condition is not of this
type. It is an explicit violation, not an anomalous violation. For this
reason it cannot be moved to the simple trace condition by means of finite
local counterterms. There is no way to enforce the conservation condition in
general, at least for the quantum field theories actually known. It is
eventually possible in new theories that admit consistent higher spin
couplings, but this is not our main concern here (we plan to develop this
issue in a future publication \cite{new}).

Off-criticality there are also anomalous violations and indeed these can be
moved to the simple tracelessness condition. We will study some of these
aspects in more depth in the next section.

Another way to study the two-point function is to write it as 
\begin{equation}
<{\cal J}_{\mu _{1}\cdots \mu _{s}}^{(s)}(x)~{\cal J}_{\nu _{1}\cdots \nu
_{s}}^{(s)}(0)>=c_{s}{\prod }_{\mu _{1}\cdots \mu _{s},\nu _{1}\cdots \nu
_{s}}^{(s)}\left( {\frac{1}{|x|^{4+2h}}}\right) ,  \label{lo}
\end{equation}
where ${\prod }_{\mu _{1}\cdots \mu _{s},\nu _{1}\cdots \nu _{s}}^{(s)}$ is
a differential operator (a polynomial of degree $2s$ in derivatives,
constructed with $\partial _{\mu }$ and $\delta _{\mu \nu }$) fixed by the
required symmetries and $c_{s}$ is a numerical factor. This form is more
convenient to impose conservation, less convenient to study conformality. It
is easy to check that conditions (\ref{doubletrace}) and (\ref{conservation}%
) do not fix this correlator (i.e. the form of ${\prod }^{(s)}$) uniquely
(see below). It is necessary to impose (\ref{simpletrace}).

Now, when (\ref{simpletrace}) holds one has also the complete conservation $%
\partial _{\mu _{1}}{\cal J}_{\mu _{1}\cdots \mu _{s}}^{(s)}=0$. A
correlator satisfying complete conservation can be easily constructed via
the projector $\pi _{\mu \nu },$ 
\begin{eqnarray}
<{\cal J}_{\mu _{1}\cdots \mu _{s}}^{(s)}(x)~{\cal J}_{\nu _{1}\cdots \nu
_{s}}^{(s)}(0)>=~~~~~~~~~~~~~~~~~~~~~~~~~~~~~~~~~~~~~~~~~~~~~~~~~~~~~~~~~~~~~~~~
\nonumber \\
~~~~~~~~\sum_{n=0}^{[s/2]}\left[ b_{n}\sum_{{\rm symm}}\pi _{\mu _{1}\mu
_{2}}\cdots \pi _{\mu _{n-1}\mu _{n}}\pi _{\nu _{1}\nu _{2}}\cdots \pi _{\nu
_{n-1}\nu _{n}}\pi _{\mu _{n+1}\nu _{n+1}}\cdots \pi _{\mu _{s}\nu
_{s}}\right] \frac{1}{|x|^{4+2h}}.  \label{ooh}
\end{eqnarray}
One has $[s/2]+1$ free parameters $b_{n}$ (different from the $a_{n}$'s) and
the anomalous dimension $h$. We now prove that (\ref{simpletrace}) imposes $%
[s/2]$ conditions and so leaves with a single overall constant, plus the
anomalous dimension $h$.

Indeed, after contraction of, say, $\mu _{s-1}$ and $\mu _{s},$ the result
has necessarily the form (using $\pi \pi =-\Box \pi $) 
\begin{equation}
\Box \sum \pi _{\nu _{s-1}\nu _{s}}\sum_{n=0}^{[s/2]-1}\left[ b_{n}^{\prime
}\sum_{{\rm symm}^{\prime }}\pi _{\mu _{1}\mu _{2}}\cdots \pi _{\mu
_{n-1}\mu _{n}}\pi _{\nu _{1}\nu _{2}}\cdots \pi _{\nu _{n-1}\nu _{n}}\pi
_{\mu _{n+1}\nu _{n+1}}\cdots \pi _{\mu _{s-2}\nu _{s-2}}\right] .
\label{evid}
\end{equation}
The internal sum reproduces the projector of a spin $s-2$ current and the
external sum takes care of the remaining permutations. The coefficients $%
b_{n}^{\prime },[s/2]$ in total, are linear combinations of the coefficients 
$b_{n}$. Therefore setting (\ref{evid}) to zero imposes the desired number
of conditions.

In this case, $h$ measures the explicit violation of conformal invariance.
Indeed, the final amplitude does not satisfy (\ref{con}) only because of $h$
and enforcing (\ref{con}) sets $h$ to 0.

This concludes the present subsection. With respect to conformality in its
general formulation (\ref{con}), the simple trace condition simplifies
enormously the search for the projector $\prod_{\mu _{1}\cdots \mu _{s}\nu
_{1}\cdots \nu _{s}}^{(s)}$ and the construction of ${\cal J}_{\mu
_{1}\cdots \mu _{s}}^{(s)}$. After presenting other general properties, we
proceed to construct the quantities (currents and correlators) for $s=3,$ $4$
and $5$.

\subsection{Orthonormality.}

Some orthonormality properties are noticeable. First we observe that for $%
s\neq s^{\prime }$ we have the orthogonality relationship 
\begin{equation}
<{\cal J}_{\mu _{1}\cdots \mu _{s}}^{(s)}(x)~{\cal J}_{\nu _{1}\cdots \nu
_{s^{\prime }}}^{(s^{\prime })}(0)>={0,}  \label{ics}
\end{equation}
at criticality. The simple trace condition is here crucial, since otherwise
we would get a nonzero result. Indeed, we shall show in the next section
that if the simple trace condition is relaxed it is always possible to
construct higher spin currents via suitable differentiation of lower spin
ones.

(\ref{ics}) holds in presence of anomalous dimensions $h_{s}$ and $%
h_{s^{\prime }}$ and it is proved as follows. Consider the most general
conformal structure of the given two-point function, constructed like in (%
\ref{confo}) and let us assume that $s<s^{\prime }$. Then one has $[s/2]+1$
free parameters before imposing the simple trace condition (the lower spin
is the one that dictates the counting). Tracing ${\cal J}^{(s^{\prime })}$
one imposes precisely $[s/2]+1$ conditions (this is correct also for $%
s^{\prime }=s+1$ due to the integral part) therefore proving the statement.
We observe that the argument based on the dimensionality of the operators
(conformal two-point functions of operators with different dimensions
vanish) is weaker than the argument that we have given: the two currents
might have precisely the same dimension even if their spins are different.

The second property that we outline has to do with the normalization of the
currents. Writing the relevant structure of (\ref{lo}) as 
\begin{equation}
{\prod }_{\mu _{1}\cdots \mu _{s},\nu _{1}\cdots \nu _{s}}^{(s)}\left( {%
\frac{1}{x^{4}}}\right) ={\cal I}_{\mu _{1}\cdots \mu _{s},\nu _{1}\cdots
\nu _{s}}^{(s)}(x)~\frac{1}{|x|^{4+2s}}{,}  \label{def}
\end{equation}
we have relationships of the form 
\begin{equation}
\frac{1}{p_{s}^{2}}~{\cal I}_{\mu _{1}\cdots \mu _{s},\alpha _{1}\cdots
\alpha _{s}}^{(s)}(x)~{\cal I}_{\alpha _{1}\cdots \alpha _{s},\nu _{1}\cdots
\nu _{s}}^{(s)}(x)=\Im _{\mu _{1}\cdots \mu _{s},\nu _{1}\cdots \nu
_{s}}^{(s)},  \label{ortos}
\end{equation}
$p_{s}$ being normalizations factors and $\Im _{\mu _{1}\cdots \mu _{s},\nu
_{1}\cdots \nu _{s}}^{(s)}$ denoting the spin-$s$ identity operators, i.e.
the identity operator on the space of symmetric, completely traceless $s$%
-indexed tensors. For example, $\Im _{\mu \nu ,\rho \sigma }^{(2)}=\frac{1}{4%
}\left[ 2\left( \delta _{\mu \rho }\delta _{\nu \sigma }+\delta _{\mu \sigma
}\delta _{\nu \rho }\right) -\delta _{\mu \nu }\delta _{\rho \sigma }\right] 
$. Property (\ref{ortos}) means that the symmetric, dimensionless matrices $%
{\cal I}_{\mu _{1}\cdots \mu _{s},\alpha _{1}\cdots \alpha _{s}}^{(s)}(x)$
are orthogonal. It generalizes the orthogonality property of the matrix $%
{\cal I}_{\mu \nu }(x)$ appearing in the Jacobian $\partial x_{\mu }^{\prime
}/\partial x_{\nu }={\cal I}_{\mu \nu }(x)/|x|^{2}=\left( \delta _{\mu \nu
}-2x_{\mu }x_{\nu }/|x|^{2}\right) /|x|^{2}$ of the coordinate inversion $%
x_{\mu }\rightarrow x_{\mu }/|x|^{2}$. We have indeed already used ${\cal I}%
_{\mu \nu }(x)$ - see formula (\ref{confo}) - to build the most general
conformal two-point correlator.

${\cal I}_{\mu \nu }(x)$ appears in the numerator of the spin-1 two-point
function (\ref{11}). For spin-2, see formula (\ref{barba2}), definition (\ref
{def}) gives ${\cal I}_{\mu \nu ,\alpha \beta }^{(2)}(x)=80(2{\cal I}_{\mu
\rho }{\cal I}_{\nu \sigma }+2{\cal I}_{\mu \sigma }{\cal I}_{\nu \rho
}-\delta _{\mu \nu }\delta _{\rho \sigma })$ and (\ref{ortos}) reads 
\begin{equation}
\frac{1}{(2^{6}\cdot 5)^{2}}~{\cal I}_{\mu \nu ,\alpha \beta }^{(2)}(x)~%
{\cal I}_{\alpha \beta ,\rho \sigma }^{(2)}(x)=\Im _{\mu \nu ,\rho \sigma
}^{(2)}.  \label{orto2}
\end{equation}
The spin 3, 4 and 5 versions of (\ref{ortos}) will be reported in the
respective subsections.

The third observation is that a cross-trace of the spin-$s$ projector ${%
\prod }_{\mu _{1}\cdots \mu _{s},\nu _{1}\cdots \nu _{s}}^{(s)}$ is a spin-$%
(s-1)$ projector. We fix the overall coefficient of ${\prod }_{\mu
_{1}\cdots \mu _{s}}^{(s)}$ so that the term $\sum_s\pi_{\mu_1\nu_1}\cdots%
\pi_{\mu_s\nu_s}$ has unit coefficient. With this convention 
\begin{equation}
{\prod }_{\mu _{1}\cdots \mu _{s-1}\alpha ,\nu _{1}\cdots \nu _{s-1}\alpha
}^{(s)}=-{\frac{2s+1}{2s-1}}\Box {\prod }_{\mu _{1}\cdots \mu _{s-1},\nu
_{1}\cdots \nu _{s-1}}^{(s-1)}.  \label{tss}
\end{equation}
The simplest case is 
\[
{\prod }_{\mu \alpha ,\alpha \rho }^{(2)}=-{\frac{5}{3}}~\Box {\pi }_{{\mu
\rho }}. 
\]
Finally, it is obvious that the square of the projector ${\prod }^{(s)}$ is
equal to itself. Precisely, the same conventions that give (\ref{tss}) give
also 
\begin{equation}
{\prod }_{\mu _{1}\cdots \mu _{s},\rho _{1}\cdots \rho _{s}}^{(s)}{\prod }%
_{\rho _{1}\cdots \rho _{s},\nu _{1}\cdots \nu _{s}}^{(s)}=(-1)^s\Box ^{s}{%
\prod }_{\mu _{1}\cdots \mu _{s},\nu _{1}\cdots \nu _{s}}^{(s)}.  \label{qs}
\end{equation}
For example, 
\[
\pi _{\mu \rho }\pi _{\rho \nu }=-\Box \pi _{\mu \nu },\quad \quad {\prod }%
_{\mu \nu ,\alpha \beta }^{(2)}{\prod }_{\alpha \beta ,\rho \sigma
}^{(2)}=\Box ^{2}{\prod }_{\mu \nu ,\rho \sigma }^{(2)} 
\]
are immediately verifiable.

\subsection{Spin 3.}

In this and in the next two subsections we analyze in detail the cases of
spin 3, 4 and 5, that will be relevant for the applications.

For $s=3$ formula (\ref{confo}) becomes 
\begin{equation}
<{\cal J}_{\mu \nu \rho }^{(3)}(x)~{\cal J}_{\alpha \beta \gamma }^{(3)}(0)>=%
\frac{1}{|x|^{10+2h}}\sum_{{\rm symm}}\left[ a_{1}\delta _{\mu \nu }\delta
_{\alpha \beta }{\cal I}_{\rho \gamma }(x)+a_{0}{\cal I}_{\mu \alpha }(x)%
{\cal I}_{\nu \beta }(x){\cal I}_{\rho \gamma }(x)\right] ,  \label{uno}
\end{equation}
with $[s/2]+1=2$ free parameters, as expected. Conservation (\ref
{conservation}) imposes $a_{0}=-2a_{1}$ and $h=0$. The simple trace
condition (\ref{simpletrace}) imposes only $a_{0}=-2a_{1}$.

Similarly, (\ref{ooh}) becomes (condition (\ref{doubletrace}) is empty for $%
s=3$) 
\begin{equation}
<{\cal J}_{\mu \nu \rho }^{(3)}(x)~{\cal J}_{\alpha \beta \gamma
}^{(3)}(0)>=\sum_{{\rm symm}}\left[ b_{1}\pi _{\mu \nu }\pi _{\alpha \beta
}\pi _{\rho \gamma }+b_{0}\pi _{\mu \alpha }\pi _{\nu \beta }\pi _{\rho
\gamma }\right] \frac{1}{|x|^{4+2h}}.  \label{due}
\end{equation}
One can check that this correlator is conformal if and only if $%
3b_{0}+5b_{1}=0$ and $h=0$. Then the simple trace condition also holds. Vice
versa, if the correlator is simply traceless and $h=0,$ then it is conserved.

At $h=0,$ when $a_{0}=-2a_{1}$ and $3b_{0}+5b_{1}=0$ expressions (\ref{uno})
and (\ref{due}) are proportional to each other. We write the final
expression of the spin 3 projector as 
\[
{\prod }_{\mu \nu \rho ,\alpha \beta \gamma }^{(3)}=\sum_{{\rm symm}}\left[
\pi _{\mu \alpha }\pi _{\nu \beta }\pi _{\rho \gamma }-\frac{3}{5}\pi _{\mu
\nu }\pi _{\alpha \beta }\pi _{\rho \gamma }\right] . 
\]
The orthonormality property (\ref{ortos}) reads 
\begin{equation}
\frac{1}{(2^{8}\cdot 3^{2}\cdot 7)^{2}}~{\cal I}_{\mu \nu \rho ,\tau \zeta
\varsigma }^{(3)}(x)~{\cal I}_{\tau \zeta \varsigma ,\alpha \beta \gamma
}^{(3)}(x)=\Im _{\mu \nu \rho ,\alpha \beta \gamma }^{(3)}.  \label{uno3}
\end{equation}
and formulas (\ref{tss})-(\ref{qs}) hold with our conventions.

We now work out the free-field expressions of the spin-3 currents. For a
scalar field (which has to be complex, as in the case of the spin-1 current)
one has 
\begin{equation}
{\cal J}_{\mu \nu \rho }^{(3)}=C^{ij}\sum_{{\rm symm}}\left( 3\partial _{\mu
}\varphi ^{i}\partial _{\nu }\partial _{\rho }\varphi ^{j}+\delta _{\mu \nu
}\partial _{\alpha }\partial _{\rho }\varphi ^{i}\partial _{\alpha }\varphi
^{j}-\frac{1}{3}\varphi ^{i}\partial _{\mu }\partial _{\nu }\partial _{\rho
}\varphi ^{j}\right) .  \label{curre}
\end{equation}
This current is uniquely fixed by the additional condition (\ref{simpletrace}%
). $C^{ij}$ is an antisymmetric matrix ($=\varepsilon ^{ij}$ for a complex
scalar in real notation). The simple trace 
\begin{equation}
{\cal J}_{\mu \mu \rho }^{(3)}=C^{ij}\left[ \partial _{\rho }\varphi
^{i}\Box \varphi ^{j}-\frac{1}{3}\varphi ^{i}\partial _{\rho }\Box \varphi
^{j}\right] ,  \label{shell}
\end{equation}
vanishes on shell.

Equivalently, the improved current (\ref{curre}) can be fixed by imposing
conservation, double tracelessness and the requirement that ${\cal J}^{(s)}$
transforms correctly under coordinate inversion, see (\ref{trasfo}). Using $%
\varphi (x)\rightarrow |x|^{2}\varphi (x),$ $\bar{\varphi}(x)\rightarrow
|x|^{2}\bar{\varphi}(x)$ and $\partial _{\mu }\rightarrow |x|^{2}{\cal I}%
_{\mu \nu }(x)\partial _{\nu }$, one can check that, indeed, (\ref{curre})
satisfies (\ref{trasfo}) and that this would not be true without the
improvement term. Similar checks can be repeated for the other currents that
appear in this and in the next subsections, but we will not mention this
fact anymore. A\ good exercise is to perform the check in the simplest
cases, namely $s=1,2,$ formulas (\ref{11}) and (\ref{12}).

For a spinor we have 
\[
{\cal J}_{\mu \nu \rho }^{(3)}=\sum_{{\rm symm}}\left[ \bar{\psi}\gamma
_{\mu }\overleftrightarrow{\partial _{\nu }}\overleftrightarrow{\partial
_{\rho }}\psi -\frac{1}{5}\pi _{\mu \nu }\left( \bar{\psi}\gamma _{\rho
}\psi \right) \right] . 
\]
The second term is the improvement term and it is constructed via the spin-1
current. There is also an axial spin-3 current 
\[
{\cal A}_{\mu \nu \rho }^{(3)}=\sum_{{\rm symm}}\left[ \bar{\psi}\gamma
_{5}\gamma _{\mu }\overleftrightarrow{\partial _{\nu }}\overleftrightarrow{%
\partial _{\rho }}\psi -\frac{1}{5}\pi _{\mu \nu }\left( \bar{\psi}\gamma
_{5}\gamma _{\rho }\psi \right) \right] , 
\]
that will be useful in the applications.

For free vectors there is no issue of improvement and the spin-3 current is
axial: 
\begin{equation}
{\cal A}_{\mu \nu \rho }^{(3)}=\frac{1}{3}\left[ F_{\nu \alpha }^{+}%
\overleftrightarrow{\partial _{\mu }}F_{\alpha \rho }^{-}+F_{\mu \alpha }^{+}%
\overleftrightarrow{\partial _{\nu }}F_{\alpha \rho }^{-}+F_{\mu \alpha }^{+}%
\overleftrightarrow{\partial _{\rho }}F_{\alpha \nu }^{-}\right] ,
\label{v3}
\end{equation}
where $F_{\mu \nu }^{\pm }=\frac{1}{2}~\left( F_{\mu \nu }\pm {\frac{1}{2}}%
\varepsilon _{\mu \nu \rho \sigma }F_{\rho \sigma }\right) ,$ $F_{\mu \nu
}=F_{\mu \nu }^{+}+F_{\mu \nu }^{-}$.

\subsection{Spin 4.}

One can repeat the same steps in the case of spin 4. (\ref{confo}) contains $%
[s/2]+1=3$ free parameters and the anomalous dimension $h$. After imposing
the simple trace condition only $h$ and the overall constant survive.
Precisely, 
\begin{eqnarray*}
<{\cal J}_{\mu \nu \rho \sigma }^{(4)}(x)~{\cal J}_{\alpha \beta \gamma
\delta }^{(4)}(0)> &=&\frac{a}{|x|^{14+2h}}\sum_{{\rm symm}}\left[ \delta
_{\mu \nu }\delta _{\rho \sigma }\delta _{\alpha \beta }\delta _{\gamma
\delta }-12\delta _{\mu \nu }\delta _{\alpha \beta }J_{\rho \gamma
}(x)J_{\sigma \delta }(x)\right. \\
&&\left. -36J_{\mu \alpha }(x)J_{\nu \beta }(x)J_{\rho \gamma }(x)J_{\sigma
\delta }(x)\right] .
\end{eqnarray*}
Conservation holds for $h=0.$ Formula (\ref{ooh}) produces the desired
projector ${\prod }^{(4)}$, after imposing the simple trace condition: 
\begin{equation}
{\prod }_{\mu \nu \rho \sigma ,\alpha \beta \gamma \delta }^{(4)}=\sum_{{\rm %
symm}}\left[ \pi _{\mu \alpha }\pi _{\nu \beta }\pi _{\rho \gamma }\pi
_{\sigma \delta }-\frac{6}{7}\pi _{\mu \nu }\pi _{\alpha \beta }\pi _{\rho
\gamma }\pi _{\sigma \delta }+\frac{3}{35}\pi _{\mu \nu }\pi _{\rho \sigma
}\pi _{\alpha \beta }\pi _{\gamma \delta }\right].  \label{spin4}
\end{equation}
The orthonormality property (\ref{ortos}) reads 
\begin{equation}
\frac{1}{(2^{14}\cdot 3^{4})^{2}}~{\cal I}_{\mu \nu \rho \sigma ,\zeta
\varsigma \xi \kappa }^{(4)}(x)~{\cal I}_{\zeta \varsigma \xi \kappa ,\alpha
\beta \gamma \delta }^{(4)}(x)=\Im _{\mu \nu \rho \sigma ,\alpha \beta
\gamma \delta }^{(4)}  \label{uno4}
\end{equation}
and (\ref{tss})-(\ref{qs}) hold.

For free scalar fields we have 
\begin{eqnarray*}
{\cal J}_{\mu \nu \rho \sigma }^{(4)} &=&{\frac{4}{35}} \sum_{{\rm symm}%
}4\varphi \partial _{\mu }\partial _{\nu }\partial _{\rho }\partial _{\sigma
}\varphi -64\partial _{\mu }\varphi \partial _{\nu }\partial _{\rho
}\partial _{\sigma }\varphi +72\partial _{\mu }\partial _{\nu }\varphi
\partial _{\rho }\partial _{\sigma }\varphi \\
&&+12\delta _{\mu \nu }\left( 2\partial _{\alpha }\varphi \partial _{\alpha
}\partial _{\rho }\partial _{\sigma }\varphi -3\partial _{\alpha }\partial
_{\rho }\varphi \partial _{\alpha }\partial _{\sigma }\varphi \right)
+3\delta _{\mu \nu }\delta _{\rho \sigma }\partial _{\alpha }\partial
_{\beta }\varphi \partial _{\alpha }\partial _{\beta }\varphi.
\end{eqnarray*}
We can write the current in a more instructive, although less explicit,
notation, 
\begin{eqnarray*}
{\cal J}_{\mu \nu \rho \sigma }^{(4)} &=& \varphi \overleftrightarrow{%
\partial _{\mu }}\overleftrightarrow{\partial _{\nu }}\overleftrightarrow{%
\partial _{\rho }}\overleftrightarrow{\partial _{\sigma }}\varphi -\frac{1}{%
24}(\delta _{\mu \nu }\delta _{\rho \sigma }+\delta _{\mu \rho }\delta _{\nu
\sigma }+\delta _{\mu \sigma }\delta _{\nu \rho })\Box ^{2}\left( \varphi
^{2}\right) \\
&&-\frac{6}{7}\sum_{{\rm symm}}\pi _{\mu \nu }\left[ \varphi 
\overleftrightarrow{\partial _{\rho }}\overleftrightarrow{\partial _{\sigma }%
}\varphi -\frac{1}{3}\pi _{\rho \sigma }\left( \varphi ^{2}\right) \right] -%
\frac{1}{5}P_{\mu \nu \rho \sigma }^{(4)}\left( \varphi ^{2}\right) .
\end{eqnarray*}
Let us explain the meaning of the various terms. The first term is the basic
term for the construction of the current, since it certainly has a spin-4
content. However, it is not purely spin-4. One has to subtract the double
trace (something which is done by the second term of the first line) and,
more importantly, one has to subtract spurious lower spin terms. This is the
role of the second line, which contains the improvement terms and enforces
the simple trace condition. In the rest of the section we shall find several
objects like these. In particular, the projector that we called $P_{\mu \nu
\rho \sigma }^{(4)}$ is the unique forth-order polynomial in derivatives
that satisfies (\ref{conservation}) and (\ref{doubletrace}) identically (but
not the simple trace condition). Its explicit expression appears in the next
section, formula (\ref{impro3}), and it subtracts an unwanted spin-0
content. Observe that the first improvement term of the second line contains
the stress-tensor $-\frac{1}{4}\left[ \varphi \overleftrightarrow{\partial
_{\rho }}\overleftrightarrow{\partial _{\sigma }}\varphi -\frac{1}{3}\pi
_{\rho \sigma }\left( \varphi ^{2}\right) \right] $ and therefore subtracts
an undesired spin-2 content.

For a spinor, 
\[
{\cal J}_{\mu \nu \rho \sigma }^{(4)}=\sum_{{\rm symm}}\left[ \bar{\psi}%
\gamma _{\mu }\overleftrightarrow{\partial _{\nu }}\overleftrightarrow{%
\partial _{\rho }}\overleftrightarrow{\partial _{\sigma }}\psi -\frac{3}{7}%
\pi _{\mu \nu }\left( \bar{\psi}\gamma _{\rho }\overleftrightarrow{\partial
_{\sigma }}\psi \right) \right] 
\]
and the improvement term is purely spin-2.

For a vector, 
\begin{equation}
{\cal J}_{\mu \nu \rho \sigma }^{(4)}=\sum_{{\rm symm}}\left[ F_{\rho \alpha
}^{+}\overleftrightarrow{\partial _{\mu }}\overleftrightarrow{\partial _{\nu
}}F_{\alpha \sigma }^{-}-\frac{1}{7}\pi _{\mu \nu }\left( F_{\rho \alpha
}^{+}F_{\alpha \sigma }^{-}\right) \right] ,  \label{j4}
\end{equation}
the improvement term is again spin-2.

\subsection{Spin 5.}

For the applications, we need also the expression of the fermionic spin-5
axial current, 
\begin{equation}
{\cal A}_{\mu \nu \rho \sigma \tau }^{(5)}=\sum_{{\rm symm}}\left[ \bar{\psi}%
\gamma _{5}\gamma _{\mu }\overleftrightarrow{\partial _{\nu }}%
\overleftrightarrow{\partial _{\rho }}\overleftrightarrow{\partial _{\sigma }%
}\overleftrightarrow{\partial _{\tau }}\psi -\frac{2}{3}\pi _{\mu \nu }{\cal %
A}_{\rho \sigma \tau }^{(3)}-\frac{3}{35}\pi _{\mu \nu }\pi _{\rho \sigma }%
{\cal A}_{\tau }^{(1)}\right] ,  \label{spin5}
\end{equation}
and the same for the vector field, 
\begin{equation}
{\cal A}_{\mu \nu \rho \sigma \tau }^{(5)}=\sum_{{\rm symm}}\left[ F_{\mu
\alpha }^{+}\overleftrightarrow{\partial _{\rho }}\overleftrightarrow{%
\partial _{\sigma }}\overleftrightarrow{\partial _{\tau }}F_{\alpha \nu
}^{-}-\frac{1}{3}\pi _{\mu \nu }\left( F_{\rho \alpha }^{+}%
\overleftrightarrow{\partial _{\tau }}F_{\alpha \sigma }^{-}\right) \right] .
\end{equation}
We see that there are spin-3 and spin-1 improvement terms. The spin-5
projector operator is

\[
{\prod }_{\mu \nu \rho \sigma \tau ,\alpha \beta \gamma \delta \varepsilon
}^{(5)}=\sum_{{\rm symm}}\left[ \pi _{\mu \alpha }\pi _{\nu \beta }\pi
_{\rho \gamma }\pi _{\sigma \delta }-\frac{10}{9}\pi _{\mu \nu }\pi _{\alpha
\beta }\pi _{\rho \gamma }\pi _{\sigma \delta }+\frac{5}{21}\pi _{\mu \nu
}\pi _{\rho \sigma }\pi _{\alpha \beta }\pi _{\gamma \delta }\right] \pi
_{\tau \varepsilon }. 
\]
It satisfies 
\[
\frac{1}{(2^{16}\cdot 3^{2}\cdot 5^{2}\cdot 11)^{2}}~{\cal I}_{\mu \nu \rho
\sigma \tau ,\zeta \varsigma \xi \kappa \iota }^{(5)}(x)~{\cal I}_{\zeta
\varsigma \xi \kappa \iota ,\alpha \beta \gamma \delta \varepsilon
}^{(5)}(x)=\Im _{\mu \nu \rho \sigma \tau ,\alpha \beta \gamma \delta
\varepsilon }^{(5)} 
\]
and (\ref{tss})-(\ref{qs}) have been verified.

\section{Critical and off-critical properties.}

In this section we study the general structure of higher spin two-point
functions. In the first subsections we assume that the higher spin
conservation law holds in nontrivially interacting quantum field theories
subject to a renormalization group flow. This is equivalent to say that the
theory in question has a higher spin flavor symmetry and it is a quite
nontrivial requirement. As we have already remarked in section \ref{ulk} and
will discuss in more detail in section 4 this does not happen in ordinary
renormalizable quantum field theories, where an explicit $h$-violation of
the conservation condition is always present. However, the investigation of
the present section is necessary as a preliminary step in order to proceed
further.

In the last subsection we include the effect of the explicit $h$-violation.
This result will be used in section 5 to define appropriate central charges
and central functions. To anticipate the idea, we remark that something
similar has been already done in ref. \cite{ccfis}. In absence of explicit
violations of the conservation condition (\ref{conservation}) the
current-current two-point functions already define the desired higher spin
central charges: this is studied in the present section and corresponds to
the work done in section 2 of \cite{ccfis} for spin-2. Instead, in presence
of explicit violations one has to construct the higher spin central charges
by studying appropriate channels of the stress-tensor four-point function.
These channels contain all possible higher spin currents. The construction
of higher spin central charges is then similar to the procedure described in
section 3 of ref. \cite{ccfis} for the spin-0 channel.

It is clear therefore, that it is convenient to start the analysis of the
current-current two-point functions under the assumption that conservation
and double tracelessness are preserved off-criticality. We also observe, and
this should not be underestimated, that if consistent higher spin couplings
do exist in some theories \cite{vasiliev}, this material will apply
straightforwardly to those theories. Alternatively, it could constitute the
basis for an axiomatic definition of these new theories.

We recall \cite{ccfis} that $<TT>$ contains two terms off-criticality, 
\begin{equation}
<T_{\mu \nu }(x)\,T_{\rho \sigma }(0)>={\frac{1}{480\pi ^{4}}\prod }_{\mu
\nu ,\rho \sigma }^{(2)}\left( {\frac{c(g(x))}{x^{4}}}\right) +\pi _{\mu \nu
}\pi _{\rho \sigma }\left( {\frac{f(\ln x\mu ,g))}{x^{4}}}\right) .
\label{two}
\end{equation}
This formula will be generalized to arbitrary spin. Via a detailed algebraic
analysis we classify the projectors that appear in the higher spin two-point
functions. We do it explicitly for spin-3 and spin-4 and then generalize the
result to spin $s$. Then, we study the correlators from the point of view of
the renormalization group in a generic renormalizable quantum field theory
with higher spin flavor symmetries, as well as anomalies and the $h$%
-violation.

The higher spin flavor symmetry imposes (\ref{doubletrace}) and (\ref
{conservation}), while (\ref{simpletrace}) holds only at criticality.
Therefore, off-criticality the current-current two-point function must
contain more invariants. We observe that trivial higher spin currents can be
constructs by suitably differentiating lower-spin currents. For example, the
operator 
\begin{equation}
\Delta _{\mu \nu \rho }^{(3,1)}=\pi _{\mu \nu }J_{\rho }+\pi _{\mu \rho
}J_{\nu }+\pi _{\nu \rho }J_{\mu }-\frac{1}{2}\left( \delta _{\mu \nu
}\partial _{\rho }+\delta _{\mu \rho }\partial _{\nu }+\delta _{\nu \rho
}\partial _{\mu }\right) \partial _{\alpha }J_{\alpha }  \label{impro}
\end{equation}
satisfies (\ref{conservation}) identically (which means that no use of the
equations of motion is necessary), for{\it \ any }spin-1 current $J_{\mu }$,
not necessarily conserved ($J_{\mu }$ could be anomalous, like the axial
current in QED, or even classically non-conserved, like the Konishi current 
\cite{konishi} in supersymmetric theories with a superpotential). This
property is precisely what defines it to be an ``improvement term''.
Therefore, (\ref{impro}) is not a genuine spin-3 current. It is more
appropriate to consider it as a higher-spin descendant of a lower-spin
current. The simple trace condition (\ref{simpletrace}) is not satisfied by $%
\Delta _{\mu \nu \rho }^{(3,1)}$.

It is natural to expect that there is some sort of conflict between ${\cal J}%
_{\mu \nu \rho }^{(3)}$ and $\Delta _{\mu \nu \rho }^{(3,1)}$. From the
point of view of renormalization, this is nothing but the operator mixing.
Let us assume, for simplicity, that $J_{\mu }$ is conserved.\ There is no
loss of generality in this, since our purpose is just to classify the
possible invariants. We write, from (\ref{11}), 
\begin{equation}
<J_{\mu }(x)~J_{\nu }(0)>=\pi _{\mu \nu }\left( \frac{c_{1}[g(t)]}{|x|^{4}}%
\right) .  \label{b}
\end{equation}
The flavor central charge $c_{1}[g(t)]$ has well-defined non-vanishing UV\
and IR\ limits \cite{noi} and the correlator is conformal in these two
limits. Then we have 
\begin{equation}
<\Delta _{\mu \nu \rho }^{(3,1)}(x)~\Delta _{\alpha \beta \gamma
}^{(3,1)}(0)>=9\sum_{{\rm symm}}\pi _{{\mu \nu }}\pi _{\rho \gamma }\pi
_{\alpha \beta }\left( \frac{c_{1}[g(t)]}{|x|^{4}}\right) .  \label{c2}
\end{equation}
This correlator is non-vanishing and not conformal in the UV\ and IR\
limits, but it is related in a simple way to a conformal correlator, via the
combined action of a certain number of derivatives on (\ref{b}). As a
further check of the mixing between spin 3 and spin 1 currents, we observe
that (\ref{conservation}) and $\partial _{\mu }J_{\mu }=0$ allow for the
correlator 
\begin{equation}
<J_{\alpha }(x)~{\cal J}_{\mu \nu \rho }^{(3)}(0)>=(\pi _{\mu \nu }\pi
_{\rho \alpha }+\pi _{\mu \rho }\pi _{\nu \alpha }+\pi _{\nu \rho }\pi _{\mu
\alpha })\left( \frac{f_{31}(t)}{|x|^{4}}\right)  \label{c3}
\end{equation}
to be nonzero, in violation of the orthogonality formula (\ref{ics}).

We are now ready to write the two-point function off-criticality. One finds
three independent invariants satisfying (\ref{conservation}), namely 
\begin{eqnarray}
<{\cal J}_{\mu \nu \rho }^{(3)}(x)~{\cal J}_{\alpha \beta \gamma
}^{(3)}(0)>= &&{\prod }_{\mu \nu \rho ,\alpha \beta \gamma }^{(3)}\left( {%
\frac{c_{3}(t)}{x^{4}}}\right) +\sum_{{\rm symm}}\pi _{{\mu \nu }}\pi _{\rho
\gamma }\pi _{\alpha \beta }\left( {\frac{c_{3,1}(t)}{x^{4}}}\right) 
\nonumber \\
&&{+{P}_{{\mu \nu \rho }}^{(3)}{P}_{{\alpha \beta \gamma }}^{(3)}}%
_{{}}\left( {\frac{c_{3,0}(t)}{x^{4}}}\right) .  \label{offc}
\end{eqnarray}
For the moment, we do not make the RG dependences of the functions $c$
explicit (in particular, do they depend only on the running coupling
constant or not?). This issue will be one of our concerns later on.

The first term of (\ref{offc}) is the pure spin-3 term, the middle term
gives the spin-3-spin-1 mixing, coming from (\ref{c2}), or, equivalently, (%
\ref{c3}). The third term is the spin3-spin-0 mixing. It is the only
factorizable invariant. Indeed the differential operator 
\[
{P}_{{\mu \nu \rho }}^{(3)}=\partial _{\mu }\partial _{\nu }\partial _{\rho
}-\frac{1}{2}\Box \left( \delta _{\mu \nu }\partial _{\rho }+\delta _{\mu
\rho }\partial _{\nu }+\delta _{\nu \rho }\partial _{\mu }\right) 
\]
satisfies (\ref{conservation}) identically. We have explicitly checked that
the three invariants in (\ref{offc}) exhaust all the possibilities.

(\ref{offc}) should be compared to the stress tensor two-point function,
formula (\ref{two}), which exhibits two invariants. One is the pure spin-2
invariant ${\prod }_{\mu \nu ,\rho \sigma }^{(2)}$ and the other is the
factorized invariant $\pi _{\mu \nu }\pi _{\rho \sigma }$ describing the
spin-2-spin-0 mixing. The improvement term $\pi _{\mu \nu }(\varphi
^{i}\varphi ^{i})$ is the analogue of (\ref{impro}).

Despite the renormalization mixing between ${\cal J}_{\mu \nu \rho }^{(3)}$
and its own improvement terms, one can isolate the function $c_{3}(t)$ by
formally extracting the traceless part via a nonlocal projection, namely 
\begin{eqnarray}
\hat{{\cal {J}}}_{\mu \nu \rho }^{(3)} &\equiv &{\cal J}_{\mu \nu \rho
}^{(3)}+\frac{1}{5\Box ^{2}}\left[ \Box \left( \pi _{\mu \nu }{\cal J}%
_{\alpha \alpha \rho }^{(3)}+\pi _{\rho \nu }{\cal J}_{\alpha \alpha \mu
}^{(3)}+\pi _{\mu \rho }{\cal J}_{\alpha \alpha \nu }^{(3)}\right) \right. 
\nonumber \\
&&\left. -\frac{1}{4}\Box \left( \delta _{\mu \nu }\partial _{\rho }+\delta
_{\mu \rho }\partial _{\nu }+\delta _{\nu \rho }\partial _{\mu }\right)
\partial _{\beta }{\cal J}_{\alpha \alpha \beta }^{(3)}-\frac{1}{2}\partial
_{\mu }\partial _{\nu }\partial _{\rho }\partial _{\beta }{\cal J}_{\alpha
\alpha \beta }^{(3)}\right] .  \label{projection}
\end{eqnarray}
Indeed, $\hat{{\cal {J}}}_{\mu \nu \rho }^{(3)}$ satisfies both (\ref
{conservation}) and (\ref{simpletrace}). It should be compared with the
nonlocal traceless projection of the stress-energy tensor (see for example
formula (1.8) of \cite{ccfis}). Therefore, 
\[
<\hat{{\cal {J}}}_{\mu \nu \rho }^{(3)}(x)~\hat{{\cal {J}}}_{\alpha \beta
\gamma }^{(3)}(0)>={\prod }_{\mu \nu \rho ,\alpha \beta \gamma }^{(3)}\left( 
{\frac{c_{3}(t)}{x^{4}}}\right) . 
\]
For the hypothetical theories with spin-3 flavor symmetry, this construction
is sufficient to isolate the desired central function $c_{3}(t)$, that will
depend only on the running coupling constant.

Before proceeding, it is useful to repeat the above analysis for spin-4
currents.

In terms of a symmetric tensor $T_{\mu \nu }$ (not necessarily traceless and
not necessarily conserved) we can always construct a spin-4 operator that
satisfies (\ref{doubletrace}) and (\ref{conservation}) identically, but not (%
\ref{simpletrace}): 
\begin{eqnarray}
\Delta _{\mu \nu \rho \sigma }^{(4,2)} &=&\pi _{\mu \nu }T_{\rho \sigma
}+\pi _{\mu \rho }T_{\nu \sigma }+\pi _{\mu \sigma }T_{\nu \rho }+\pi _{\rho
\sigma }T_{\mu \nu }+\pi _{\nu \sigma }T_{\mu \rho }+\pi _{\nu \rho }T_{\mu
\sigma } \\
&&-\frac{1}{2}(\delta _{\mu \nu }\partial _{\rho }\partial _{\alpha
}T_{\alpha \sigma }+{\rm perms}_{12})+\,{\frac{1}{12}}\,(\delta _{\mu \nu
}\delta _{\rho \sigma }+\delta _{\mu \rho }\delta _{\nu \sigma }+\delta
_{\mu \sigma }\delta _{\nu \rho })\left( 5\Box T+4\partial _{\alpha
}\partial _{\beta }T_{\alpha \beta }\right) ,  \nonumber
\end{eqnarray}
where $T=T_{\mu \nu }$. For clarity we have specified via a subscript the
total number of terms that one obtains after permutation (we shall use the
same convention in other cases). Note that several improvement terms found
in section 2 are of this type.

Secondly, one can construct an object with the same properties by means of
the spin-1 current $J_{\mu }$, namely 
\begin{eqnarray}
\Delta _{\mu \nu \rho \sigma }^{(4,1)} &=&P_{\mu \nu \rho }^{(3)}J_{\sigma
}+P_{\nu \rho \sigma }^{(3)}J_{\mu }+P_{\rho \sigma \mu }^{(3)}J_{\nu
}+P_{\sigma \mu \nu }^{(3)}J_{\rho }-\frac{1}{3}(\delta _{\mu \nu }\partial
_{\rho }\partial _{\sigma }+{\rm perms}_{6})\partial _{\alpha }J_{\alpha }
\label{impro2} \\
&&+\frac{1}{2}(\delta _{\mu \nu }\delta _{\rho \sigma }+\delta _{\mu \rho
}\delta _{\nu \sigma }+\delta _{\mu \sigma }\delta _{\nu \rho })\Box
\partial _{\alpha }J_{\alpha }.  \nonumber
\end{eqnarray}
Finally, the only factorizable invariant is generated by the forth-order
differential operator 
\begin{eqnarray}
P_{\mu \nu \rho \sigma }^{(4)} &=&\partial _{\mu }\partial _{\nu }\partial
_{\rho }\partial _{\sigma }-{\frac{1}{3}}\Box (\partial _{\mu }\partial
_{\nu }\delta _{\rho \sigma }+\partial _{\mu }\partial _{\rho }\delta _{\nu
\sigma }+\partial _{\mu }\partial _{\sigma }\delta _{\nu \rho }+\partial
_{\nu }\partial _{\rho }\delta _{\mu \sigma }+\partial _{\nu }\partial
_{\sigma }\delta _{\mu \rho }+\partial _{\rho }\partial _{\sigma }\delta
_{\mu \nu })  \nonumber \\
&&+{\frac{1}{8}}\Box ^{2}(\delta _{\mu \nu }\delta _{\rho \sigma }+\delta
_{\mu \rho }\delta _{\nu \sigma }+\delta _{\mu \sigma }\delta _{\nu \rho }),
\label{impro3}
\end{eqnarray}
the unique one that satisfies identically (\ref{doubletrace}) and (\ref
{conservation}) and represents the mixing between spin 4 and spin zero. For
example, $\Delta _{\mu \nu \rho \sigma }^{(4,0)}=P_{\mu \nu \rho \sigma
}^{(4)}[\varphi ^{2}]$ is a spin-0 local operator mixing with ${\cal J}_{\mu
\nu \rho \sigma }^{(4)}$, $\varphi $ being a scalar field.

Off-criticality, the two-point function $<{\cal J}^{(4)}{\cal J}^{(4)}>$
contains four invariants.

i) One invariant is the pure spin-4 invariant ${\prod }^{(4)}$ of (\ref
{spin4}).

ii) A spin-2 invariant comes from $<\Delta ^{(4,2)}\Delta ^{(4,2)}>$ and has
the form $\pi {\prod }^{(2)}\pi $;

iii) $<\Delta ^{(4,1)}\Delta ^{(4,1)}>$ produces a spin-1 invariant the form 
$P^{(3)}\pi P^{(3)}$.

iv) Finally, there is the spin-0 invariant $\left( P^{(4)}\right) ^{2}$.

\noindent Correlators of the type $<\Delta ^{(4,2)}\Delta ^{(4,1)}>$, $<%
{\cal J}^{(4)}\Delta ^{(4,1)}>$ and so on, do not produce anything new. For
example, $<T_{\mu \nu }T_{\rho \sigma }>$ contains also a spin-0 invariant
of the form $\pi \pi $, as we know. Therefore, $<\Delta ^{(4,2)}\Delta
^{(4,2)}>$ contributes itself a spin-0 invariant of the form $\pi \pi \pi
\pi +\pi \pi \Box ^{2}\delta \delta $. One can verify that this spin-0
invariant is the same as $\left( P^{(4)}\right) ^{2}$.

The claimed four invariants are clearly independent. We have carefully
checked that they exhaust all possibilities. In summary, the most general
spin-4 two-point function in a spin-4 flavor symmetric theory is

\begin{eqnarray}
<{\cal J}_{\mu \nu \rho \sigma }^{(4)}(x)~{\cal J}_{\alpha \beta \gamma
\delta }^{(4)}(0)> &=&{\prod }_{\mu \nu \rho \sigma ,\alpha \beta \gamma
\delta }^{(4)}\left( {\frac{c_{4}(t)}{x^{4}}}\right) +\sum_{{\rm symm}}\pi _{%
{\mu \nu }}{\prod }_{\rho \sigma ,\alpha \beta }^{(2)}\pi _{\gamma \delta
}\left( {\frac{c_{4,2}(t)}{x^{4}}}\right)  \nonumber \\
&+&\sum_{{\rm symm}}P_{\mu \nu \rho }^{(3)}\pi _{\sigma \alpha }P_{\beta
\gamma \delta }^{(3)}\left( {\frac{c_{4,1}(t)}{x^{4}}}\right) \ {+\ {P}_{{%
\mu \nu \rho \sigma }}^{(4)}{P}_{{\alpha \beta \gamma \delta }}^{(4)}}%
_{{}}\left( {\frac{c_{4,0}(t)}{x^{4}}}\right) .  \label{44}
\end{eqnarray}
These and other formulas in our paper might appear rather implicit, but
after some experience one easily recognizes the structure of the terms. For
example, the fourth term of (\ref{44}) is completely factorized. The third
and second term are partially factorized (which means that each term in the
sum is factorized, but the sum itself is not factorized) - in particular,
the third term in (\ref{44}) is, so to speak, 3/4-factorized, while the
second term is 1/2-factorized. Finally, the first term exhibits no
factorization. These structures reflect the spin content of each invariant.

It is important to show that there exists a unique nonlocal projection
isolating the pure spin-4 central function $c_{4}(t)$ from the mixing
functions $c_{4,i}(t)$, $i\leq s-2$, like in (\ref{projection}). The result
is 
\begin{eqnarray}
\hat{{\cal {J}}}_{\mu \nu \rho \sigma }^{(4)} &=&{\cal J}_{\mu \nu \rho
\sigma }^{(4)}+{\frac{1}{7}}{\frac{1}{\Box }}[\pi _{\mu \nu }{\cal J}%
_{\alpha \alpha \rho \sigma }^{(4)}+{\rm perms}_{6}]  \nonumber \\
&&-{\frac{2}{21}}{\frac{1}{\Box ^{2}}}\left[ \partial _{\mu }\partial _{\nu
}\partial _{\rho }\partial _{\alpha }{\cal J}_{\sigma \alpha \beta \beta
}^{(4)}+{\rm perms}_{4}+{\frac{1}{4}}\Box \delta _{\mu \nu }\partial _{\rho
}\partial _{\alpha }{\cal J}_{\sigma \alpha \beta \beta }^{(4)}+{\rm perms}%
_{12}\right]  \nonumber \\
&&+{\frac{2}{105}}{\frac{1}{\Box ^{2}}}\left[ \delta _{\mu \nu }\partial
_{\rho }\partial _{\sigma }+{\rm perms}_{6}+{\frac{1}{4}}\Box (\delta _{\mu
\nu }\delta _{\rho \sigma }+\delta _{\mu \rho }\delta _{\nu \sigma }+\delta
_{\mu \sigma }\delta _{\nu \rho })\right] \partial _{\alpha }\partial
_{\beta }{\cal J}_{\alpha \beta \gamma \gamma }^{(4)}  \nonumber \\
&&+{\frac{4}{105}}{\frac{1}{\Box ^{3}}}\partial _{\mu }\partial _{\nu
}\partial _{\rho }\partial _{\sigma }\partial _{\alpha }\partial _{\beta }%
{\cal J}_{\alpha \beta \gamma \gamma }^{(4)}.  \label{projection4}
\end{eqnarray}
Finally, we can throw away all sorts of complicacies by writing the
projected two-point function 
\[
<\hat{{\cal {J}}}_{\mu \nu \rho \sigma }^{(4)}(x)~\hat{{\cal {J}}}_{\alpha
\beta \gamma \delta }^{(4)}(0)>={\prod }_{\mu \nu \rho \sigma ,\alpha \beta
\gamma \delta }^{(4)}\left( {\frac{c_{4}(t)}{x^{4}}}\right) , 
\]
which is what we wanted to arrive at.

The analysis can be easily generalized to spin $s$. We present here the
result. The most general two-point function has the form 
\begin{equation}
<{\cal J}_{\mu _{1}\cdots \mu _{s}}^{(s)}(x)~{\cal J}_{\nu _{1}\cdots \nu
_{s}}^{(s)}(0)>={\prod }_{\mu _{1}\cdots \mu _{s},\nu _{1}\cdots \nu
_{s}}^{(s)}\left( {\frac{c_{s}(t)}{x^{4}}}\right) +\sum_{s^{\prime }=0}^{s-2}%
{\prod }_{\mu _{1}\cdots \mu _{s},\nu _{1}\cdots \nu _{s}}^{(s,s^{\prime
})}\left( {\frac{c_{s,s^{\prime }}(t)}{x^{4}}}\right) ,  \label{auto}
\end{equation}
where the sum runs over the spin mixings. There is no $s$-$\left( s-1\right) 
$ spin mixing and a unique $s$-$s^{\prime }$ spin mixing for any $s^{\prime
}=s-2,\ldots ,0$. There exists a unique nonlocal projection $\hat{{\cal {J}}}%
^{(s)}$ isolating the pure spin-$s$ part, such that 
\begin{equation}
<\hat{{\cal {J}}}_{\mu _{1}\cdots \mu _{s}}^{(s)}(x)~\hat{{\cal {J}}}_{\nu
_{1}\cdots \nu _{s}}^{(s)}(0)>={\prod }_{\mu _{1}\cdots \mu _{s},\nu
_{1}\cdots \nu _{s}}^{(s)}\left( {\frac{c_{s}(t)}{x^{4}}}\right) .
\label{nonlo}
\end{equation}
At criticality $\hat{{\cal {J}}}_{\mu _{1}\cdots \mu _{s}}^{(s)}={\cal J}%
_{\mu _{1}\cdots \mu _{s}}^{(s)}$ and $c_{s}(t)=c_{s*}$ is a constant. (\ref
{nonlo}) reduces then to (\ref{lo}).

\subsection{Renormalization group analysis.}

The relationship between bare and renormalized operators mixes the higher
spin current with all its improvement terms and is therefore of the form 
\begin{equation}
{\cal J}_{\mu _{1}\cdots \mu _{s}}^{(s)R}={\cal J}_{\mu _{1}\cdots \mu
_{s}}^{(s)B}+\sum_{s^{\prime }=0}^{s-2}\ A_{s,s^{\prime }}\ \Delta _{\mu
_{1}\cdots \mu _{s}}^{(s,s^{\prime })B}.  \label{rg}
\end{equation}
We recall that we are still assuming that the spin-$s$ tensor current ${\cal %
J}_{\mu _{1}\cdots \mu _{s}}^{(s)}$ satisfies (\ref{doubletrace}) and (\ref
{conservation}) on shell and the above formula follows from this assumption.

In particular, the coefficient in front of ${\cal J}_{\mu _{1}\cdots \mu
_{s}}^{(s)B}$ is one. Indeed, conditions (\ref{doubletrace}) and (\ref
{conservation}) are assumed to be violated by terms proportional to the
field equations: terms of this type are finite operators, this being a
well-known classical result (see for example the book by Collins \cite
{collins}). Therefore, the renormalization constants of the operators 
\[
{{{{{{\cal J}^{(s)}}_{\mu }}^{\mu }}_{\nu }}^{\nu }}_{\mu _{5}\cdots \mu
_{s}} 
\]
and 
\[
\partial ^{\mu }{\cal J}_{\mu \mu _{2}\cdots \mu _{s}}^{(s)}-{\frac{1}{%
2(s-1)(s-2)}}\sum_{i<j=2}^{s}\delta _{\mu _{i}\mu _{j}}\partial ^{\mu }{%
{\cal J}^{(s)}}_{\mu \cdots \mu _{i-1}\alpha \mu _{i+1}\cdots \mu
_{j-1}\alpha \mu _{j+1}\cdots \mu _{s}} 
\]
equal one, as we wanted to prove. This does not imply that ${\cal J}_{\mu
_{1}\cdots \mu _{s}}^{(s)}$ is itself finite, since (\ref{doubletrace}) and (%
\ref{conservation}) are identically satisfied (i.e. satisfied off-shell) by
certain operators that we call $\Delta _{\mu _{1}\cdots \mu
_{s}}^{(s,s^{\prime })}$. The operators $\Delta _{\mu _{1}\cdots \mu
_{s}}^{(s,s^{\prime })}$ are constructed iteratively via the lower spin
tensor currents ${\cal J}_{\mu _{1}\cdots \mu s^{\prime }}^{(s^{\prime })}$, 
$s^{\prime }=s-2,\ldots ,0,$ (or in general non-conserved lower spin
operators) by acting with derivatives in various ways, like in the formulas
that we have worked out explicitly for $s=3$ and $s=4,$ (\ref{impro}), (\ref
{impro2}), (\ref{impro3}) and so on. Moreover, as we have proved, $\Delta
_{\mu _{1}\cdots \mu _{s}}^{(s,s^{\prime })}$ are the only operators with
this property and therefore they renormalize among themselves. $%
A_{s,s^{\prime }}$ are the appropriate renormalization mixing constants. It
follows from our assumptions that the set of operators formed by ${\cal J}%
_{\mu _{1}\cdots \mu _{s}}^{(s)B}$ and $\Delta _{\mu _{1}\cdots \mu
_{s}}^{(s,s^{\prime })B}$, $s^{\prime }=s-2,\ldots ,0,$ is closed under
renormalization mixing.

Equations (\ref{rg}) assure that the $s\times s$ renormalization matrix $%
Z_{ij}$, $i,j=s,s-2,\ldots ,0,$ of the set of operators $\left( {\cal J}%
_{\mu _{1}\cdots \mu _{s}}^{(s)B},\Delta _{\mu _{1}\cdots \mu
_{s}}^{(s,s^{\prime })B}\right) $ is triangular and $Z_{11}=1$. Then the
analysis of section 1 of ref. \cite{ccfis} can be repeated straightforwardly
to prove that the functions $c_{s}$ depend just on the running coupling
constant, $c_{s}(t)=c_{s}[g(t)],$ and therefore are good central functions
under the assumption of higher spin flavor symmetry.

\subsection{Trace anomaly and higher spin anomalies.}

Taking the scale derivative $\mu ~\partial /\partial \mu $ of (\ref{lo}) one
gets the expression for the integrated trace anomaly. The integrated trace
anomaly reads at criticality

\begin{equation}
\int \Theta =\sum_{s=1}^{\infty }c_{s*}\int \ h_{s}^{\mu _{1}\cdots \mu _{s}}%
{\prod }_{\mu _{1}\cdots \mu _{s},\nu _{1}\cdots \nu _{s}}^{(s)}h_{s}^{\nu
_{1}\cdots \nu _{s}}+{\cal O}(h^{3}).  \label{ac}
\end{equation}
Note that the $c_{s}$-term is the higher spin analogue of the square of the
Weyl tensor for gravity. It is the conformal term of the higher derivative
action for higher spin fields. At intermediate energies one has instead 
\begin{equation}
\int \Theta =\sum_{s=1}^{\infty }\int h_{s}^{\mu _{1}\cdots \mu _{s}}\left[ 
\tilde{c}_{s}(g){\prod }_{\mu _{1}\cdots \mu _{s},\nu _{1}\cdots \nu
_{s}}^{(s)}+\sum_{s^{\prime }=0}^{s-2}\tilde{c}_{s,s^{\prime }}(g){\prod }%
_{\mu _{1}\cdots \mu _{s},\nu _{1}\cdots \nu _{s}}^{(s,s^{\prime })}\right]
h_{s}^{\nu _{1}\cdots \nu _{s}}+{\cal O}(h^{3}).
\end{equation}
The relationship between the tilded functions of this formula to the
untilded ones of the previous subsections can be found in section 1 of ref. 
\cite{ccfis}. The complete higher derivative action for higher spin fields
is also encoded in (\ref{auto}) and reads: 
\begin{equation}
{\cal S}_{s}=\int h_{s}^{\mu _{1}\cdots \mu _{s}}\left[ \alpha _{s}{\prod }%
_{\mu _{1}\cdots \mu _{s},\nu _{1}\cdots \nu _{s}}^{(s)}+\sum_{s^{\prime
}=0}^{s-2}\alpha _{s,s^{\prime }}{\prod }_{\mu _{1}\cdots \mu _{s},\nu
_{1}\cdots \nu _{s}}^{(s,s^{\prime })}\right] h_{s}^{\nu _{1}\cdots \nu
_{s}}+{\cal O}(h^{3}),  \label{act}
\end{equation}
$\alpha _{s},\alpha _{s,s^{\prime }}$ being the coupling constants. The
higher orders in $h$ should be determined by the gauge invariance (\ref
{gaugeinv}) (itself suitably modified by higher orders in $h$) and are
encoded in the divergent parts of the $n$-point correlators 
\begin{equation}
<{\prod_{i=1}^{n}}{\cal J}_{\mu _{1}^{i}\cdots \mu _{s}^{i}}^{(s)}(x_{i})>.
\label{one}
\end{equation}
The higher derivative higher spin action is expected to be power-counting
renormalizable and therefore generalize the familiar higher derivative
action for quantum gravity, 
\[
\int \sqrt{g}\left( \alpha R^{2}+\beta R_{\mu \nu }R^{\mu \nu }\right) , 
\]
to which the lower derivative action ($\int \sqrt{g}R$ for gravity, the
action of ref.s \cite{fronsdal,dewit} for higher spin fields) can be added
without spoiling renormalizability.

The unintegrated trace anomaly can contain topological terms, at least for
even $s$. In particular at criticality we have 
\begin{eqnarray}
\Theta &=&\sum_{s=1}^{\infty }c_{s*}\ h_{s}^{\mu _{1}\cdots \mu _{s}}{\prod }%
_{\mu _{1}\cdots \mu _{s},\nu _{1}\cdots \nu _{s}}^{(s)}h_{s}^{\nu
_{1}\cdots \nu _{s}}  \label{topo} \\
&&+\sum_{s={\rm even}}^{\infty }a_{s*}\ \partial _{\mu _{1}\cdots \mu
_{s}}\left[ h_{s}^{\nu _{1}\cdots \nu _{s}}\partial _{\rho _{1}\cdots \rho
_{s}}h_{s}^{\sigma _{1}\cdots \sigma _{s}}\right] {\prod_{i=1}^{s}\ }\left( {%
\varepsilon }_{\mu _{i}\nu _{i}\rho _{i}\sigma _{i}}\right) +{\cal O}(h^{3})
\nonumber
\end{eqnarray}
where $\partial _{\mu _{1}\cdots \mu _{s}}=\partial _{\mu _{1}}\cdots
\partial _{\mu _{s}}.$ Off-criticality the topological invariant is
multiplied by a function $a_{s}(g)$. Therefore the complete unintegrated
trace anomaly contains $s+1$ terms in total, those of (\ref{act}) plus the
topological invariant appearing in (\ref{topo}). For $s$ odd the topological
invariant that we have written in (\ref{topo}) violates parity and appears
in the axial anomalies.

The ones that we have described are all the allowed higher derivative
invariants, as we now prove. Summarizing, they are $s+1$ in total, one being
topological, one being conformal, the other $s-1$ invariants disappearing at
criticality.

To complete the proof that we have listed all possible invariants for the
trace anomaly (the analysis of the previous subsection suffices to prove
this fact up to total derivatives), we need to show that there exists no
other independent topological term. We can proceed as follows. Let us recall
that there exists a ``Riemann'' tensor for higher spin fields. To our
knowledge, the first expression of this tensor appeared in \cite{dewit}.
Nevertheless, here we write the form of ref. \cite{berends}, more explicit
for our purpose, namely 
\[
R_{\mu _{1}\nu _{1}\cdots \mu _{s}\nu _{s}}=\partial _{\alpha _{1}\cdots
\alpha _{s}}h_{s}^{\beta _{1}\cdots \beta _{s}}\prod_{i=1}^{s}(\delta _{\mu
_{i}\alpha _{i}}^{{}}\delta _{\nu _{i}\beta _{i}-}\delta _{\mu _{i}\beta
_{i}}^{{}}\delta _{\nu _{i}\alpha _{i}}). 
\]
We can introduce the ``Weyl'' tensor by extracting all traces from $R_{\mu
_{1}\nu _{1}\cdots \mu _{s}\nu _{s}}$ and we can define $s$ ``Ricci''
tensors by tracing $R_{\mu _{1}\nu _{1}\cdots \mu _{s}\nu _{s}}$ in all
possible ways. In total, one counts $s+1$ independent tensors. Squaring
these tensors one reproduces exactly $s+1$ higher derivative terms. These
are the $s$ terms of the higher derivative action (\ref{act}) and the
topological term contained in (\ref{topo}). Indeed, expanding the
topological term of (\ref{topo}) by using the properties of the $\varepsilon 
$-tensor, one can re-express it as a linear combination of the squares of
the ``Weyl'' tensor and the $s$ ``Ricci'' tensors (this is the
``Gauss-Bonnet'' theorem for higher spin fields).

We do not re-derive our classification of invariants following this
procedure, since we have already obtained it in a different manner. We leave
this completion as an exercise for the reader. It is enough to note that the
counting proves that all $s+1$ possibilities have been accounted for, as
claimed. Topological terms linear in the curvature (like $\Box R$ for
spin-2) are not listed.

We conclude that there exist precisely one central charge of type $c_{s}$
and one central charge of type $a_{s}$ for each spin $s$. The quantities $%
a_{s}$ are expected to satisfy irreversibility properties (``$a$-theorems'')
analogous to the property satisfied by the coefficient of the Euler density
in the expression of the trace anomaly in external gravity (the quantity $%
a_{2}$ \cite{noi}). These $a$-theorems are expressed by inequalities $%
a_{sUV}-a_{sIR}\geq 0.$

The properties of the quantities $a_{s}$ will be studied in detail in a
forthcoming publication \cite{new}, where other higher spin anomalies will
also be treated quantitatively. We observe, in this context, that the simple
trace ${\cal J}_{\mu _{1}\cdots \mu _{s-2}\alpha \alpha }^{(s)}$ of a spin-$%
s $ current ${\cal J}^{(s)}$ (the role played by $\Theta $ for the
stress-tensor) is a dimension $2+s$ simply traceless operator, not
conserved, with $s-2$ indices and therefore $(s-1)^{2}$ components in total.
It can be decomposed into a sum 
\begin{equation}
{\cal J}_{\mu _{1}\cdots \mu _{s-2}\alpha \alpha }^{(s)}=\sum_{s^{\prime
}=0}^{s-2}\Theta _{\mu _{1}\cdots \mu _{s-2}}^{(s,s^{\prime })},
\label{deco}
\end{equation}
each term being simply traceless and having a definite spin-$s^{\prime }$
content. With this we mean that $\Theta _{\mu _{1}\cdots \mu
_{s-2}}^{(s,s^{\prime })}$ is an operator with the same quantum numbers as
the traces $\Delta _{\mu _{1}\cdots \mu _{s-2\alpha \alpha }}^{(s,s^{\prime
})}$ of the improvement operators. For example, for $s=2$, we have $%
s^{\prime }=0$ and $\Theta $ has indeed the same quantum numbers as the
trace of $\pi _{\mu \nu }\varphi ^{2}$, the unique improvement term for the
stress-tensor. Similarly, the two-point functions $<\Theta _{\mu _{1}\cdots
\mu _{s-2}}^{(s,s^{\prime })}(x)~\Theta _{\nu _{1}\cdots \nu
_{s-2}}^{(s,s^{\prime })}(0)>$ are of the same form as $<\Delta _{\mu
_{1}\cdots \mu _{s-2\alpha \alpha }}^{(s,s^{\prime })}(x)~\Delta _{\mu
_{1}\cdots \mu _{s-2\beta \beta }}^{(s,s^{\prime })}(0)>.$

Decomposition (\ref{deco}) can be proved by observing that taking the simple
trace of the two-point function (\ref{auto}) one remains only with the
claimed mixing terms. Each of them describes the mixing with a unique lower
spin operator, with spin $s^{\prime }$ between 0 and $s-2.$ The counting
indeed matches:\ $2s^{\prime }+1$ components for each $s^{\prime }$ make in
total $\sum_{s^{\prime }=0}^{s-2}(2s^{\prime }+1)=(s-1)^{2}$ components.

This concludes the discussion of higher spin currents in theories with
higher spin flavor symmetries. Higher spin currents, however, are useful
tools to study certain aspects of ordinary renormalizable quantum field
theories, where no flavor higher spin symmetry is present, and indeed the
major purpose of the paper is to improve our knowledge of such aspects. The
results of the present section will have to be combined with other results
in order to achieve this goal.

\subsection{Treatment of the $h$-violation.}

The $h$-violation is responsible for one additional term in formula (\ref
{auto}). We perform a simple counting to explain this fact. We know that a
symmetric $k$-indexed tensor $S_{\mu _{1},\ldots \mu _{k}}$ has a number of
components equal to the number of decompositions of the integer $k$ into the
sum of four integer numbers, $k=n_{1}+n_{2}+n_{3}+n_{4}$. This is the number
of ways to assign the values 1,2,3,4 to the indices $\mu _{1},\ldots \mu
_{k} $ : there will be $n_{1}$ $1$'s, $n_{2}$ $2$'s, and so on. It follows
from a simple calculation that a symmetric traceless $k$-indexed tensor has $%
(k+1)^{2}$ components. A doubly traceless $s$-indexed tensor ${\cal J}^{(s)}$
is the sum of an $s$-indexed traceless tensor and an $(s-2)$-indexed
traceless tensor and therefore has $(s+1)^{2}+(s-1)^{2}=2(s^{2}+1)$
components, which we can write as 
\[
2(s^{2}+1)=(s+1)^{2}+(s-1)^{2}=\sum_{s^{\prime }=0}^{s}(2s^{\prime
}+1)+\sum_{s^{\prime }=0}^{s-2}(2s^{\prime }+1), 
\]
in order to exhibit the spin decomposition. The divergence operator (\ref
{conservation}) is an $(s-1)$-indexed traceless tensor, therefore it has $%
s^{2}$ components. The difference is 
\begin{equation}
s^{2}+2=(2s+1)+\sum_{s^{\prime }=0}^{s-2}(2s^{\prime }+1).  \label{more}
\end{equation}
This is precisely the spin decomposition of formula (\ref{auto}): a spin-$s$
term, with projector ${\prod }^{(s)}$, plus mixing terms of the type ($%
s,s^{\prime }$) for $s^{\prime }=s-2,\ldots 0$, with projectors ${\prod }%
^{(s,s^{\prime })}$.

So far we have re-described the results of the previous subsections. Now we
proceed differently. We do not impose the conservation condition, which
indeed does not hold in general interacting theories, but we nevertheless
enforce the simple trace condition (\ref{simpletrace}). We can always do
this since this condition is purely algebraic. We remain with 
\begin{equation}
(s+1)^{2}=\sum_{s^{\prime }=0}^{s}(2s^{\prime
}+1)=(2s+1)+(2s-1)+\sum_{s^{\prime }=0}^{s-2}(2s^{\prime }+1)  \label{less}
\end{equation}
components. We have one component more than in (\ref{more}), precisely the
spin $s-1$ term, for which we could not find a projector ${\prod }^{(s,s-1)}$%
. Now we understand the deep reason of this fact: if such a projector
existed then we would have automatically conservation.

(\ref{auto}) will contain precisely one additional term. We observe that two
different types of violations of the conservation condition appear in (\ref
{less}): the terms of the sum $\sum_{s^{\prime }=0}^{s-2}(2s^{\prime }+1)$
are the violations that we call {\it anomalous}, since they can be moved to
the simple trace condition (via finite local counterterms) and be described
by the projectors ${\prod }^{(s,s^{\prime })}$; the term $(2s-1)$ is the 
{\it explicit} violation, responsible for the anomalous dimension $h$. This
term is not anomalous, because it cannot be moved to the simple trace
condition, since there exists no projector ${\prod }^{(s,s-1)}$.

Anomalous terms are moved from the conservation condition to the trace
condition or vice versa by the simple redefinition 
\[
{\cal J}_{\mu _{1}\cdots \mu _{s}}^{(s)}\rightarrow {\cal J}_{\mu _{1}\cdots
\mu _{s}}^{(s)}+\sum_{{\rm symm}}\delta _{\mu _{s-1}\mu _{s}}\Theta _{\mu
_{1}\cdots \mu _{s-2}}, 
\]
where $\Theta _{\mu _{1}\cdots \mu _{s-2}}$ is traceless (compare this with
formula (\ref{deco})). It is clear that these additions cannot have spin
higher than $s-2$ and accordingly there is no possibility to cancel the $h$%
-violation.

So, let us move the anomalous violations to the simple trace condition. The
terms $(2s+1)$ and $(2s-1)$ can be described by a sum of the right hand
sides of (\ref{lo}) and (\ref{def}). We relabel the functions and conclude
that the most general two-point correlator reads 
\begin{equation}
<{\cal J}_{\mu _{1}\cdots \mu _{s}}^{(s)}(x)~{\cal J}_{\nu _{1}\cdots \nu
_{s}}^{(s)}(0)>={\frac{c_{s}(t)}{|x|^{4+2s}}}{\cal I}_{\mu _{1}\cdots \mu
_{s},\nu _{1}\cdots \nu _{s}}^{(s)}(x)+\sum_{s^{\prime }=0}^{s-1}{\prod }%
_{\mu _{1}\cdots \mu _{s},\nu _{1}\cdots \nu _{s}}^{(s,s^{\prime })}\left( {%
\frac{c_{s,s^{\prime }}(t)}{x^{4}}}\right) .  \label{auto2}
\end{equation}
Here ${\prod }^{(s,s-1)}$ is defined to be ${\prod }^{(s)},$ with a certain
abuse of notation. The first term (\ref{auto2}) is the one that survives at
criticality: 
\[
c_{s}(t)\sim \frac{c_{s*}}{|x|^{2h_{s}}}. 
\]
The other functions vanish at criticality. Precisely, this means 
\begin{equation}
\frac{c_{s,s^{\prime }}(t)}{c_{s}(t)}\rightarrow 0,~~~~~~~~~s^{\prime
}=s-1,\ldots ,0.  \label{prop}
\end{equation}
If $h_{s}=0$ (free fixed point) then ${\prod }^{(s)}\left( 1/|x|^{4}\right) $
is the same as ${\cal I}^{(s)}(x)/|x|^{4}$. In this case the function $%
c_{s,s-1}$ is {\it defined} to satisfy property (\ref{prop}), in order to
avoid overcounting. This is the sense in which $c_{s,s-1}$ is ``declassed''
to describe the spin-$s$-spin-$(s-1)$ mixing.

In order to familiarize with (\ref{auto2}), let us recall that in the
simplest case, $s=1$, formula (\ref{auto2}) becomes 
\[
<{\cal J}_{\mu }(x)~{\cal J}_{\nu }(0)>=c_{1}(t)\frac{{\cal I}_{\mu \nu }(x)%
}{|x|^{6}}+{\pi }_{\mu \nu }\left( {\frac{c_{1,0}(t)}{x^{4}}}\right) , 
\]
which can be understood as a decomposition into the sum of the ``conformal''
part and the ``conserved'' part of the correlator. This formula can be used,
for example, for the axial-current two-point function.

One of the by-products of the investigation of the present paper is that
higher spin tensor currents and the axial current fall in the same class.
Indeed, axial currents and higher spin currents have various properties in
common: the violations of their conservation conditions vanish only in the
free fixed point; nonvanishing anomalous dimensions survive in the
interacting fixed points (see \cite{noialtri}); they appear in the same
context, in particular in the $TT$ operator product expansion (see next
section).

We remark that the usual anomalous violation of the axial-current
conservation, $\partial_\mu j_5^\mu=-{\frac{e^2}{16\pi^2}}F\tilde F, $ is
called in our language ``explicit violation'', since it does not vanish at a
generic fixed point, but only at a free fixed point. Instead, the usual
trace anomaly of the stress-tensor, $\Theta=-{\frac{\beta}{4\alpha}}F^2, $
is of the type that we call ``anomalous'', since it vanishes at both fixed
points.

We have worked out free-field formulas and classified improvement terms,
projectors, two-point functions off-criticality, as well as anomalous and
explicit violations of the trace and conservation conditions. But we have
not given the concrete expressions of the currents off-criticality. There is
a certain amount of arbitrariness (for example, how to weight the scalar,
vector and spinor contributions?). We will see in the next sections that the
relevant higher spin currents are defined by the $TT$ OPE, but before
proceeding further, let us make some final observation.

One would like to define an interpolating (``primary'') current such that
its UV and IR limits are conformal, which amounts to separate the good
current from its own improvement terms. Indeed, these are not conformal
operators, but descendants of conformal operators. For example, a
redefinition like 
\begin{equation}
{\cal J}_{\mu _{1}\cdots \mu _{s}}^{(s)}\rightarrow {\cal J}_{\mu _{1}\cdots
\mu _{s}}^{(s)}+\sum_{s^{\prime }=0}^{s-2}\ h_{s,s^{\prime }}(g)\ \Delta
_{\mu _{1}\cdots \mu _{s}}^{(s,s^{\prime })},  \label{ir}
\end{equation}
where $h_{s,s^{\prime }}(g)$ are arbitrary functions of the coupling
constants, finite and vanishing in the free-field limit ($h_{s,s^{\prime
}}(0)=0)$, is always allowed. It does not change the free (UV) current,
uniquely determined in section 2. However, it does change the IR\ limit of $%
{\cal J}^{(s)}$ if $h_{s,s^{\prime }}(g_{IR})\neq 0.$ The correlators of $%
\Delta _{\mu _{1}\cdots \mu _{s}}^{(s,s^{\prime })}$ are not conformal (see
for example (\ref{c2})) and so one can spoil conformality of the two-point
function $<{\cal J}^{(s)}{\cal J}^{(s)}>$ in the IR, in particular property (%
\ref{prop}). Only the precise knowledge of the dynamics of the theory can
fix (\ref{ir}) so that it interpolates appropriately between two conformal
currents. For spin-1, on the other hand, where there is no improvement term,
the current (vector or axial) is uniquely fixed. For higher spin the
ambiguity survives even in presence of a higher spin symmetry. The stress
tensor, for example, can be redefined as $T_{\mu \nu }\rightarrow T_{\mu \nu
}+h(g)\pi _{\mu \nu }[\varphi ^{2}]$ and the trace as $\Theta \rightarrow
\Theta -3h(g)\Box \varphi ^{2}$. This corresponds to a redefinition of the
action in external gravity, ${\cal L}\rightarrow {\cal L}+h(g)R\varphi ^{2},$
via a finite counterterm. The additional vertex generates divergent
counterterms of the type $R^{2}$. As a consequence, the trace anomaly in
external gravity has a new coefficient in front of $R^{2}.$ We recall that
this coefficient is the function that we called $\tilde{c}_{2,0}(g)$ (or $%
c_{2,0}(g)$ at the level of the two-point function $<TT>)$. In conclusion,
redefinition (\ref{ir}) affects the functions $c_{s,s^{\prime }}(g)$, $%
s^{\prime }<s$, spoiling eventually (\ref{prop}), but leaving the relevant
central function $c_{s}(g)$ unchanged. This is a ``mild'' effect and can for
most purposes be neglected. One interpolates between UV\ and IR with any
preferred form (\ref{ir}) of the current. If the IR\ limit of $<{\cal J}%
^{(s)}{\cal J}^{(s)}>$ , formula (\ref{auto2}), is not conformal, one
extracts the conformal part either via the nonlocal projection ${\cal J}%
^{(s)}\rightarrow \hat{{\cal {J}}}^{(s)} $ or via the spin decomposition (%
\ref{auto2}).

\section{The $TT$ operator product expansion.}

Higher spin currents appear in the operator product expansion of the stress
energy tensor. This is true in any quantum field theory in more that two
dimensions. Only in two dimensions the $TT$ OPE closes with the identity
operator and $T$ itself, several other operators appearing in general
dimension \cite{citare}. In paper \cite{noialtri}, the first of our series,
it was stressed that this mixing can have interesting applications, in
interacting conformal field theories as well as in quantum field theory out
of the critical points. It is therefore mandatory to deepen our knowledge on
this issue. In the present section we work out the $TT$ OPE for free fields
in complete detail and use these results to write down the most general form of
the $TT$ OPE in interacting conformal field theories. In the next section we
shall combine the result of the present and previous sections to achieve our
goal and define higher spin central charges and central functions.
We focus on symmetric tensors in our analysis of the OPE.

Our results are consistent with those found in earlier works,
in particular in the context of deep inelastic scattering \cite{muta},
where the operator product light-cone expansion of two electromagnetic
currents was extensively studied. Comparison with
the results of ref.s \cite{muta,1*} makes it apparent that the ligh-cone
and Euclidean expansions are very different, in the sense that the terms
are organized in a different way. For example, infinitely many
operators, including their
descendants, have the same light-cone singularity in the free-field
limit and different singularities in the Euclidean framework.
The space-time structure of the singular terms is
correspondingly rearranged. The $TT$ OPE was 
not previously considered in detail.

\begin{equation}
\begin{tabular}{|c|c|c|c|c|c|c|c|c|c|}
\hline
{\scriptsize Singularity} & $1/x^{8}$ & $1/x^{7}$ & $1/x^{6}$ & $1/x^{5}$ & $%
1/x^{4}$ & $1/x^{3}$ & $1/x^{2}$ & $1/x$ & $1$ \\ \hline
{\scriptsize Spin} & $-$ & $-$ & $0$ & $1$ & $2$ & $3$ & $4$ & $5$ & $\cdots 
$ \\ \hline
{\scriptsize Operator} & 1 & $-$ & $\Sigma $ & ${\cal A}_{\mu }^{(1)}$ & $%
T_{\mu \nu }$ & ${\cal A}_{\mu \nu \rho }^{(3)}$ & ${\cal J}_{\mu \nu \rho
\sigma }^{(4)}$ & ${\cal A}_{\mu \nu \rho \sigma \alpha }^{(5)}$ & 
{\scriptsize regular} \\ \hline
{\scriptsize Descendants} & $-$ & $-$ & $-$ & $\partial \Sigma $ & $
\begin{array}{c}
\partial {\cal A}^{(1)} \\ 
\partial ^{2}\Sigma
\end{array}
$ & $
\begin{array}{c}
\partial T \\ 
\partial ^{2}{\cal A}^{(1)} \\ 
\partial ^{3}\Sigma
\end{array}
$ & $
\begin{array}{c}
\partial {\cal A}^{(3)} \\ 
\partial ^{2}T \\ 
\partial ^{3}{\cal A}^{(1)} \\ 
\partial ^{4}\Sigma
\end{array}
$ & $
\begin{array}{c}
\partial {\cal J}^{(4)} \\ 
\partial ^{2}{\cal A}^{(3)} \\ 
\partial ^{3}T \\ 
\partial ^{4}{\cal A}^{(1)} \\ 
\partial ^{5}\Sigma
\end{array}
$ & $\cdots $ \\ \hline
\end{tabular}
\label{TTOPE}
\end{equation}

A quick look at the singularity structure of the $TT$ OPE reveals that it
contains the operators shown in the table. In particular, there are
operators of spin from 0 to 5 before the regular terms. However, this table
does not say how many independent operators of each spin there are. To
answer this question we need to classify all possible invariants for free
scalars, spinors and vectors.

\subsection{Scalar field.}

We have the propagator (using $\Box (1/|x|^{2})=-4\pi ^{2}\delta (x)$) 
\[
<\varphi (x)~\varphi (y)>=\frac{1}{4\pi ^{2}}\frac{1}{|x-y|^{2}}, 
\]
while the stress tensor is given in formula (\ref{barba2}). The $TT$ OPE
contains currents with even spin: 0,2,4. Precisely,

\begin{eqnarray}
T_{\mu \nu }(x)~T_{\rho \sigma }(y) &=&\frac{1}{60}~\left( \frac{1}{4\pi ^{2}%
}\right) ^{2}~{\prod }_{\mu \nu ,\rho \sigma }^{(2)}\left( \frac{1}{|x-y|^{4}%
}\right)  \nonumber \\
&&+\frac{1}{6}~\frac{1}{4\pi ^{2}}~\Sigma ~{\prod }_{\mu \nu ,\rho \sigma
}^{(2)}\left( \frac{1}{|x-y|^{2}}\right)  \nonumber \\
&&+\frac{1}{4\pi ^{2}}~T_{\alpha \beta }~{\prod }_{\mu \nu ,\rho \sigma
;\alpha \beta }^{(2,2;2)}\left( |x-y|^{2}\ln |x-y|^{2}M^{2}\right)
\label{ope} \\
&&+\frac{1}{4\pi ^{2}}~{\cal J}_{\alpha \beta \gamma \delta }^{(4)}~{\prod }%
_{\mu \nu ,\rho \sigma ;\alpha \beta \gamma \delta }^{(2,2;4)}\left(
|x-y|^{6}\ln |x-y|^{2}M^{2}\right)  \nonumber \\
&&+{\rm descendants}+{\rm regular\ terms.}  \nonumber
\end{eqnarray}
The operators that appear on the right hand side of the OPE are located in
the point $(x+y)/$ $2$ (this will not be written explicitly in order to
simplify the formulas). The general term (for $s<6$, in order to be
singular) is therefore of the form 
\begin{equation}
{\cal J}_{\alpha _{1}\cdots \alpha _{s}}^{(s)}~{\prod }_{\mu \nu ,\rho
\sigma ;\alpha _{1}\cdots \alpha _{s}}^{(2,2;s)}\left[ |x-y|^{2s-2}\left(
\ln |x-y|^{2}M^{2}\right) ^{\sigma (2s-2)}\right] .  \label{prima}
\end{equation}
The logarithm is present only if $s\geq 1$: $\sigma (2s-2)=1$ for $s\geq 1$
and $\sigma (2s-2)=0$ for $s=0$. The scale $M$ disappears after taking the
derivatives and has no physical meaning. The differential operator ${\prod }%
_{\mu \nu ,\rho \sigma ;\alpha _{1}\cdots \alpha _{s}}^{(2,2;s)}$ is fixed
according to the following rules: it is of order $s+4$ in derivatives,
symmetric in $\mu \leftrightarrow \nu $, $\rho \leftrightarrow \sigma $, $%
\mu \nu \leftrightarrow \rho \sigma $ and in $\alpha _{1}\cdots \alpha _{s}$
(in total, $2^{3}s!$ exchanges); it is completely traceless in $\alpha
_{1}\cdots \alpha _{s}$, as well as in $\mu \nu $ and $\rho \sigma $;
finally, it is conserved in $\mu \nu \rho \sigma $ (but not necessarily in $%
\alpha _{1}\cdots \alpha _{s}$). A term like (\ref{prima}) will be called
``primary''.

In addition to the primary terms, we have their descendants and regular
terms. Most of this section is devoted to the primary terms and before
starting their classification, let us illustrate the descendants.

The descendant terms are uniquely fixed by the primary terms and the
symmetry constraints on the operator product expansion. For example, let us
consider the first descendant of the spin-0 operator $\Sigma =\varphi ^{2}$.
The primary term 
\begin{equation}
\frac{1}{6}\frac{1}{4\pi ^{2}}\Sigma \left( \frac{x+y}{2}\right) {\prod }%
_{\mu \nu ,\rho \sigma }^{(2)}\left( \frac{1}{|x-y|^{2}}\right)
\label{primary}
\end{equation}
is conserved only up to $\partial \Sigma $-terms. The first descendant is
fixed by requiring conservation up to $\partial ^{2}\Sigma $-terms. Let us
write the first descendant of (\ref{primary}) as 
\[
\frac{1}{6}\frac{1}{4\pi ^{2}}\partial _{\alpha }\Sigma \left( \frac{x+y}{2}%
\right) {\prod }_{\mu \nu ,\rho \sigma ;\alpha }^{({\rm des})}\left( \ln
|x-y|^{2}M^{2}\right) . 
\]
The differential operator ${\prod }_{\mu \nu ,\rho \sigma ;\alpha }^{({\rm %
des})}$ is allowed to contain five derivatives, it is traceless in $\mu \nu $
and $\rho \sigma ,$ but not conserved, rather it satisfies 
\begin{equation}
\partial _{\alpha }{\prod }_{\alpha \nu ,\rho \sigma ;\mu }^{({\rm des}%
)}\left( \ln |x-y|^{2}\mu ^{2}\right) +\frac{1}{2}{\prod }_{\mu \nu ,\rho
\sigma }^{(2)}\left( \frac{1}{|x-y|^{2}}\right) =0.  \label{un}
\end{equation}
Finally, symmetry under the exchange $(\mu \nu ,x)\leftrightarrow (\rho
\sigma ,y)$ imposes 
\begin{equation}
{\prod }_{\mu \nu ,\rho \sigma ;\alpha }^{({\rm des})}+{\prod }_{\rho \sigma
,\mu \nu ;\alpha }^{({\rm des})}=0.  \label{du}
\end{equation}
(\ref{un}) and (\ref{du}) produce the unique answer 
\begin{equation}
{\prod }_{\mu \nu ,\rho \sigma ;\alpha }^{({\rm des})}=\frac{1}{8}\left( {%
\prod }_{\mu \nu ,\rho \alpha }^{(2)}\partial _{\sigma }+{\prod }_{\mu \nu
,\alpha \sigma }^{(2)}\partial _{\rho }-{\prod }_{\mu \alpha ,\rho \sigma
}^{(2)}\partial _{\nu }-{\prod }_{\alpha \nu ,\rho \sigma }^{(2)}\partial
_{\mu }\right) .  \label{desce}
\end{equation}
This formula gives the projector of the first descendant in terms of the
projector of the primary. Similar formulas hold in all other cases, but one
(a special term SP that we shall discuss at length). We call the objects ${%
\prod }^{(2,2;s)}$ ``projector-primaries'' and ${\prod }^{({\rm des})}$
``projector-descendants''. We conclude that the projector-descendants are
generated algorithmically from the projector-primaries. Therefore, we do not
need to illustrate the descendant terms any further, our main concern being
now to classify the primary terms.

A useful property to work out our classification in the following. We
observe that the set of derivatives $\partial _{\{\mu _{1}\cdots \mu _{m}}%
{\cal J}_{\nu _{1}\cdots \nu _{n}\}}^{(n)}$, completely symmetrized in the
indices and with $m$ and $n$ even, form a basis for the quadratic monomials
of the form $\partial _{\{\mu _{1}\cdots \mu _{m}}\varphi ~\partial _{\nu
_{1}\cdots \nu _{n}\}}^{{}}\varphi $, $m+n=$even. Indeed, it is easy to show
that the correspondence between the two sets of operators is one-to-one. The
condition of simple tracelessness (\ref{simpletrace}) is important to avoid
overcounting: for example, a traceful stress tensor $T_{\mu \nu }$ can only
be proportional to $\partial _{\alpha }\varphi \partial _{\alpha }\varphi $,
but this is equal to $\frac{1}{2}\Box \Sigma $. Actually, all the terms with
contracted derivatives, like $\partial _{\alpha }\varphi \partial _{\alpha
}\varphi =\frac{1}{2}\Box \Sigma $, either vanish using the field equations
or can be expressed trivially by acting with ``$\Box $'' on simpler
operators. Similarly, a spin-4 current satisfying the double trace condition
but not the simple trace condition would have a simple trace equal to a
linear combination of $\partial _{\{\mu }\partial _{\nu }T_{\rho \sigma \}},$
$\partial _{\mu }\partial _{\nu }\partial _{\rho }\partial _{\sigma }\Sigma $
and various contractions of these descendants. This fact follows by counting
the possible terms. In the table we show the correspondence in the first few
cases.

We conclude that each monomial $\partial _{\{\mu _{1}\cdots \mu _{m}}\varphi
~\partial _{\nu _{1}\cdots \nu _{n}\}}^{{}}\varphi $ appearing on the right
hand side of the OPE can be reexpressed as the sum of a primary term of spin 
$m+n$ plus descendants of operators with lower spins. In this way, its
primary content is easily identified. However, this argument does not say
anything, yet, about the number of independent operators with the same spin
appearing in the OPE, for which we need to classify the projectors ${\prod }%
^{(2,2;s)}$.

\begin{equation}
\begin{tabular}{|c|c|}
\hline
$\varphi ^{2}$ & $\Sigma $ \\ \hline
$\partial _{\mu }\varphi \partial _{\nu }\varphi $ & $T_{\mu \nu }+\frac{1}{6%
}\partial _{\mu }\partial _{\nu }\Sigma +\frac{1}{12}\delta _{\mu \nu }\Box
\Sigma $ \\ \hline
$\varphi \partial _{\mu }\partial _{\nu }\varphi $ & $-T_{\mu \nu }+\frac{1}{%
3}\partial _{\mu }\partial _{\nu }\Sigma -\frac{1}{12}\delta _{\mu \nu }\Box
\Sigma $ \\ \hline
$\sum_{{\rm s}}\partial _{\mu }\partial _{\nu }\varphi \partial _{\rho
}\partial _{\sigma }\varphi $ & $\frac{1}{16}{\cal J}_{\mu \nu \rho \sigma
}^{(4)}+K_{\mu \nu \rho \sigma } +\frac{2}{7}H_{\mu \nu \rho \sigma }+\frac{1%
}{30}\partial _{\mu }\partial _{\nu }\partial _{\rho }\partial _{\sigma
}\Sigma $ \\ \hline
$\sum_{{\rm s}}\partial _{\mu }\varphi \partial _{\nu }\partial _{\rho
}\partial _{\sigma }\varphi $ & $-\frac{1}{16} {\cal J}_{\mu \nu \rho \sigma
}^{(4)}-K_{\mu \nu \rho \sigma } +\frac{3}{14}H_{\mu \nu \rho \sigma }+\frac{%
1}{20}\partial _{\mu }\partial _{\nu }\partial _{\rho }\partial _{\sigma
}\Sigma $ \\ \hline
$\varphi \partial _{\mu }\partial _{\nu }\partial _{\rho }\partial _{\sigma
}\varphi ~$ & $\frac{1}{16}{\cal J}_{\mu \nu \rho \sigma }^{(4)}+K_{\mu \nu
\rho \sigma } -\frac{12}{7}H_{\mu \nu \rho \sigma }+\frac{1}{5}\partial
_{\mu }\partial _{\nu }\partial _{\rho }\partial _{\sigma }\Sigma $ \\ \hline
notation: $H_{\mu \nu \rho \sigma }$ & $\sum_{{\rm s}}\left( \partial _{\mu
}\partial _{\nu }T_{\rho \sigma }+\frac{1}{12}\delta _{\mu \nu }\partial
_{\rho }\partial _{\sigma }\Box \Sigma \right) $ \\ \hline
notation: $K_{\mu \nu \rho \sigma }$ & ${\frac{3}{70}}\Box \sum_{{\rm s}%
}\delta _{\mu \nu }\left( 5T_{\rho \sigma }+\frac{1}{6}\partial _{\rho
}\partial _{\sigma }\Sigma +\frac{7}{24}\delta _{\rho \sigma }\Box \Sigma
\right) $ \\ \hline
\end{tabular}
\label{table1}
\end{equation}

Let us start with ${\prod }^{(2,2;2)}$. With six uncontracted derivatives
two obvious candidates for ${\prod }_{\mu \nu ,\rho \sigma ;\alpha \beta
}^{(2,2;2)}$ are 
\begin{equation}
{\prod }_{\mu \nu \alpha ,\beta \rho \sigma }^{(3)}+{\prod }_{\mu \nu \beta
,\alpha \rho \sigma }^{(3)}+\frac{7}{10}\delta _{\alpha \beta }\Box {\prod }%
_{\mu \nu ,\rho \sigma }^{(2)},\quad \quad {\prod }_{\mu \nu ,\rho \sigma
}^{(2)}\left( \partial _{\alpha }\partial _{\beta }-\frac{1}{4}\Box \delta
_{\alpha \beta }\right) .  \label{thetwo}
\end{equation}
In the first candidate we have used (\ref{tss}) to extract the trace. Note
that terms like the first one exhibit a spin-3 content in the three-point
function of the stress- tensor. The above projectors can be generated in a
way that we now illustrate and that generalizes straightforwardly to ${\prod 
}_{\mu \nu ,\rho \sigma ;\alpha _{1}\cdots \alpha _{s}}^{(2,2;s)}$.

We have an OPE term containing a projector ${\prod }_{\mu \nu ,\rho \sigma
;\alpha \beta }^{(2,2;2)}$ with $6$ indices. An OPE term is the two-point
limit of a three-point function, so let us regard it as a peculiar two-point
function, a two-point function whose spin content has still to be determined
and where one operator is actually the product of two operators. Certainly
in ${\prod }_{\mu \nu ,\rho \sigma ;\alpha \beta }^{(2,2;2)}$ there can be a
spin-$3$ term, whose two-point function is described by our familiar
projector ${\prod }^{(3)}$, completely symmetric in its indices. The indices 
$\mu \nu ,\rho \sigma ,\alpha \beta $ can be distributed apparently in
different manners. Actually, one can check that, after imposing the correct
symmetries and taking traces out, one remains with a single independent
choice. We have written this term in the form ${\prod }_{\mu \nu \alpha
,\beta \rho \sigma }^{(3)}$ shown in (\ref{thetwo}). This term satisfies
more properties than required (for example, it is conserved in $\alpha \beta 
$) and therefore it cannot be the unique solution. Six indices can also
describe the two-point function of a product operator of spin (2-1), and
this is the second term in (\ref{thetwo}). There is no other possibility,
since there is no other decomposition of three indices.

To show that this argument generalizes further, let us now discuss the case $%
s=4$. The first candidate is the spin-4 projector ${\prod }^{(4)}$. As
before, there is a unique independent way of distributing the indices. The
other decompositions of four indices are the products (3-1) and (2-2). In
total, three independent tensors, a fact that we have carefully checked. The
invariants are therefore 
\begin{equation}
{\prod }_{\mu \nu \rho \sigma ,\alpha \beta \gamma \delta }^{(4)},\quad \sum 
\Sb {\rm symm}  \\ (\alpha \beta \gamma \delta )  \endSb {\prod }_{\mu \nu
\alpha ,\rho \sigma \beta }^{(3)}\left( \partial _{\gamma }\partial _{\delta
}-\frac{1}{4}\Box \delta _{\gamma \delta }\right) ,\quad {\prod }_{\mu \nu
,\rho \sigma }^{(2)}\left( \partial _{\alpha }\partial _{\beta }\partial
_{\gamma }\partial _{\delta }-{\rm tr}_{(\alpha \beta \gamma \delta
)}\right) .  \label{the3}
\end{equation}

In general, one takes the total number of indices of the projector ($s+4$
for ${\prod }^{(2,2;s)}$ - we assume $s$=even for simplicity), divides it by
two $s^{\prime }=s/2$ (since a projector for a two-point function has twice
as many indices as the current) and counts the number of ways of splitting $%
s^{\prime }$ into two (since a term in the OPE is a two-point limit of a
three-point function): $s^{\prime }=m+n$, $m\leq n$.

The three invariants of (\ref{the3}) do reproduce the appropriate term in
the OPE (\ref{ope}) (see below). However, this is not true for the two terms
(\ref{thetwo}). The reason is that there is one projector more in the case
for ${\prod }_{\mu \nu ,\rho \sigma ;\alpha \beta }^{(2,2;2)},$ that we have
neglected by applying the argument outlined above.

So far we have classified the projectors by imposing the conservation
condition and the condition of tracelessness in a strict sense. However,
there is one case in which these properties can be satisfied up to local
terms, i.e. up to a delta-function. This can only happen for the singularity 
$1/|x|^{4}$: lower singularities cannot produce any $\delta ^{(4)}(x)$,
while higher singularities, which would eventually produce $\Box \delta
^{(4)}(x)$ or $\partial \delta ^{(4)}(x)$, have already been discussed in
detail. The special additional invariant for ${\prod }_{\mu \nu ,\rho \sigma
;\alpha \beta }^{(2,2;2)}$ is unique, quadratic in derivatives, acts on $%
1/|x|^{2}$ and factorizes a $\Box $ when tracing or appropriately
differentiating, 
\begin{eqnarray*}
{\rm SP}_{\mu \nu ,\rho \sigma ;\alpha \beta } &=&\frac{1}{2}\left[ (\delta
_{\mu \nu }\delta _{\rho \sigma }-\delta _{\mu \rho }\delta _{\nu \sigma
}-\delta _{\mu \sigma }\delta _{\nu \rho })\partial _{\alpha }\partial
_{\beta }\right. -2\delta _{\alpha \mu }\delta _{\beta \nu }\partial _{\rho
}\partial _{\sigma }-2\delta _{\alpha \rho }\delta _{\beta \sigma }\partial
_{\mu }\partial _{\nu } \\
&&+(\delta _{\beta \rho }\delta _{\nu \sigma }+\delta _{\beta \sigma }\delta
_{\nu \rho }-\delta _{\beta \nu }\delta _{\rho \sigma })\partial _{\alpha
}\partial _{\mu }+(\delta _{\beta \rho }\delta _{\mu \sigma }+\delta _{\beta
\sigma }\delta _{\mu \rho }-\delta _{\beta \mu }\delta _{\rho \sigma
})\partial _{\alpha }\partial _{\nu } \\
&&+(\delta _{\beta \mu }\delta _{\nu \sigma }+\delta _{\beta \nu }\delta
_{\mu \sigma }-\delta _{\beta \sigma }\delta _{\mu \nu })\partial _{\alpha
}\partial _{\rho } \\
&&\left. +(\delta _{\beta \mu }\delta _{\nu \rho }+\delta _{\beta \nu
}\delta _{\mu \rho }-\delta _{\beta \rho }\delta _{\mu \nu })\partial
_{\alpha }\partial _{\sigma }\right] +(\alpha \leftrightarrow \beta )-{\rm tr%
}_{\alpha \beta }.
\end{eqnarray*}
This term is the one which closes the Poincar\'{e} algebra (we will check
this explicitly later on).

By explicit computation, the projector ${\prod }_{\mu \nu ,\rho \sigma
;\alpha \beta }^{(2,2;2)}$ is in the case of the scalar field the following
linear combination: 
\begin{eqnarray}
&&{\prod }_{\mu \nu ,\rho \sigma ;\alpha \beta }^{(2,2;2)}\left(
|x-y|^{2}\ln |x-y|^{2}M^{2}\right) ={\rm SP}_{\mu \nu ,\rho \sigma ;\alpha
\beta }\left( \frac{1}{|x-y|^{2}}\right)  \nonumber \\
&&+\frac{1}{6}{\prod }_{\mu \nu ,\rho \sigma }^{(2)}\partial _{\alpha
}\partial _{\beta }\left( |x-y|^{2}\ln |x-y|^{2}M^{2}\right)  \nonumber \\
&&-\frac{5}{16}{\prod }_{\mu \nu \alpha ,\beta \rho \sigma }^{(3)}\left(
|x-y|^{2}\ln |x-y|^{2}M^{2}\right) .\quad \quad  \label{ft}
\end{eqnarray}
For $s=4$, instead, the projector is 
\begin{eqnarray*}
{\prod }_{\mu \nu ,\rho \sigma ;\alpha \beta \gamma \delta }^{(2,2;4)} &=&-%
\frac{1}{2^{11}\cdot 3^{3}\cdot 5}\left[ {\prod }_{\mu \nu ,\rho \sigma
}^{(2)}\left( \partial _{\alpha }\partial _{\beta }\partial _{\gamma
}\partial _{\delta }-{\rm tr}_{(\alpha \beta \gamma \delta )}\right) \right.
\\
&&\left. -\frac{75}{28}\sum\Sb {\rm symm}  \\ (\alpha \beta \gamma \delta ) 
\endSb {\prod }_{\mu \nu \alpha ,\rho \sigma \beta }^{(3)}\left( \partial
_{\gamma }\partial _{\delta }-\frac{1}{4}\Box \delta _{\gamma \delta
}\right) -\frac{15}{8}{\prod }_{\mu \nu \rho \sigma ,\alpha \beta \gamma
\delta }^{(4)}\right] .
\end{eqnarray*}
The spin-4 two-point functions is 
\[
<{\cal J}_{\mu \nu \rho \sigma }^{(4)}(x)~{\cal J}_{\alpha \beta \gamma
\delta }^{(4)}(0)>=\frac{2^4}{3^2\cdot 5\cdot 7}\left( \frac{1}{4\pi ^{2}}%
\right) ^{2}{\prod }_{\mu \nu \rho \sigma ,\alpha \beta \gamma \delta
}^{(4)}\left( {\frac{1}{|x|^{4}}}\right) , 
\]
while the spin-0 two point function is simply 
\[
<\Sigma (x)~\Sigma (0)>=\frac{1}{4\pi ^{2}}\frac{2}{|x|^{4}}. 
\]

Summarizing, there are three invariants in both cases $s=2$ and $s=4.$ The
explicit coefficients with which they appear in the operator product
expansion depend on the nature of the field (scalar, spinor, vector).
Therefore there are three spin-2 central charges, three spin-4 central
charges, and so on. This conclusion should be compared with the conclusion
of the previous section, which stated that at the level of the trace anomaly
(and under the assumption of higher spin flavor symmetry) there are only two
central charges for any spin. This is not a contradiction, since it
corresponds to two different descriptions, although not unrelated.

Before concluding this section, let us make one further remark. In the
spin-2 projector, ${\prod }^{(2,2;2)}$, the existence of the three
invariants that we have listed reflects the known fact \cite{stanev} that
the three-point function of the stress tensor contains three independent
structures, corresponding precisely to scalar, spinor and vector fields.
This can be explained as follows. Primary and descendant terms are
sufficient to reconstruct the entire three-point correlator. We know that
the descendants are uniquely fixed by the primary terms. Therefore the
three-point function is fully encoded in the primary term and the number of
structures of the three-point correlator has to be the same as the number of
structures appearing in the OPE. This idea is completely general and can be
used as an algorithm to classify all the structures of the three-point
functions, for any spin, as we have done explicitly up to $s=4$.

\subsection{Vector field.}

Let us now repeat the analysis in the case of the vector field. The
situation is somewhat simpler, because the $TT$ OPE closes up to $1/|x|^{4}$%
-terms. However, odd spin currents appear for lower singularities. We have 
\[
T_{\mu \nu }=F_{\mu \alpha }F_{\nu \alpha }-\frac{1}{4}\delta _{\mu \nu
}F^{2}=2F_{\mu \alpha }^{+}F_{\nu \alpha }^{-},\qquad <A_{\mu }(x)~A_{\nu
}(0)>=\frac{1}{4\pi ^{2}}\frac{\delta _{\mu \nu }}{|x|^{2}}.
\]
The OPE reads 
\begin{eqnarray*}
T_{\mu \nu }(x)T_{\rho \sigma }(y) &=&\frac{1}{5}~\left( \frac{1}{4\pi ^{2}}%
\right) ^{2}~{\prod }_{\mu \nu ,\rho \sigma }^{(2)}\left( \frac{1}{|x-y|^{4}}%
\right)  \\
&&\!\!\!\!\!\!\!\!\!\!\!\!\!\!\!\!\!\!\!\!\!\!\!\!\!\!\!\!\!\!\!\!\!\!\!\!\!%
\!\!\!\!\!{+\frac{1}{4\pi ^{2}}~T_{\alpha \beta }~{\rm SP}_{\mu \nu ,\rho \sigma
;\alpha \beta }\left( \frac{1}{|x-y|^{2}}\right) } \\
&&\!\!\!\!\!\!\!\!\!\!\!\!\!\!\!\!\!\!\!\!\!\!\!\!\!\!\!\!\!\!\!\!\!\!\!\!\!%
\!\!\!\!\!{-\frac{1}{4\pi ^{2}}~{\cal A}_{\alpha \beta \gamma }^{(3)}~\frac{1}{2^{3}}%
\sum_{{\rm symm}}\varepsilon _{\mu \rho \alpha \delta }~\left( \pi _{\nu
\beta }~\pi _{\sigma \gamma }-~\pi _{\nu \sigma }~\partial _{\beta }\partial
_{\gamma }\right) \partial _{\delta }{~}\left( |x-y|^{2}\ln
|x-y|^{2}M^{2}\right) } \\
&&\!\!\!\!\!\!\!\!\!\!\!\!\!\!\!\!\!\!\!\!\!\!\!\!\!\!\!\!\!\!\!\!\!\!\!\!\!%
\!\!\!\!\!{+\frac{1}{4\pi ^{2}}~{\cal J}_{\alpha \beta \gamma \delta }^{(4)}~{\prod }%
_{\mu \nu ,\rho \sigma ;\alpha \beta \gamma \delta }^{(2,2;4)}\left(
|x-y|^{6}\ln |x-y|^{2}M^{2}\right) } \\
&&\!\!\!\!\!\!\!\!\!\!\!\!\!\!\!\!\!\!\!\!\!\!\!\!\!\!\!\!\!\!\!\!\!\!\!\!\!%
\!\!\!\!\!{-\frac{1}{4\pi ^{2}}~{\cal A}_{\alpha \beta \gamma \delta \varepsilon
}^{(5)}~\frac{1}{2^{12}\cdot 3^{2}}\sum_{{\rm symm}}\varepsilon _{\mu \rho
\alpha \zeta }~~\left( \pi _{\nu \beta }~\pi _{\sigma \gamma }-~\pi _{\nu
\sigma }~\partial _{\beta }\partial _{\gamma }\right) \partial _{\delta
}\partial _{\varepsilon }\partial _{\zeta }\left( |x-y|^{6}\ln
|x-y|^{2}M^{2}\right) } \\
&&\!\!\!\!\!\!\!\!\!\!\!\!\!\!\!\!\!\!\!\!\!\!\!\!\!\!\!\!\!\!\!\!\!\!\!\!\!%
\!\!\!\!\!{+{\rm descendants}+{\rm regular\ terms.}}
\end{eqnarray*}
The odd spin terms have been written down explicitly, since there is a
unique projector for each of them (see below). We see that there is only the
special invariant SP\ for $s=2$ and that it multiplies the stress tensor
with the same coefficient as in the scalar case. The conversion table now
reads 
\begin{equation}
\begin{tabular}{|c|c|}
\hline
$F_{\mu \alpha }^{+}F_{\nu \alpha }^{-}$ & $\frac{1}{2}T_{\mu \nu }$ \\ 
\hline
$F_{\mu \nu }^{+}F_{\rho \sigma }^{-}+F_{\rho \sigma }^{+}F_{\mu \nu }^{-}$
& $\frac{1}{2}\left[ \delta _{\mu \rho }T_{\nu \sigma }-\delta _{\mu \sigma
}T_{\nu \rho }-\delta _{\nu \rho }T_{\mu \sigma }+\delta _{\nu \sigma
}T_{\mu \rho }\right] $ \\ \hline
$\sum_{{\rm s}}\left( F_{\rho \alpha }^{+}\partial _{\mu }\partial _{\nu
}F_{\sigma \alpha }^{-}+\partial _{\mu }\partial _{\nu }F_{\rho \alpha
}^{+}F_{\sigma \alpha }^{-}\right) $ & $-\frac{1}{2}{\cal J}_{\mu \nu \rho
\sigma }^{(4)}+\frac{1}{28}\sum_{{\rm s}}\pi _{\rho \sigma }T_{\mu \nu }+%
\frac{1}{4}\sum_{{\rm s}}\partial _{\rho }\partial _{\sigma }T_{\mu \nu }$
\\ \hline
$\sum_{{\rm s}}\left( \partial _{\mu }F_{\rho \alpha }^{+}\partial _{\nu
}F_{\sigma \alpha }^{-}+\partial _{\nu }F_{\rho \alpha }^{+}\partial _{\mu
}F_{\sigma \alpha }^{-}\right) $ & $\frac{1}{2}{\cal J}_{\mu \nu \rho \sigma
}^{(4)}-\frac{1}{28}\sum_{{\rm s}}\pi _{\rho \sigma }T_{\mu \nu }+\frac{1}{4}%
\sum_{{\rm s}}\partial _{\rho }\partial _{\sigma }T_{\mu \nu }$ \\ \hline
\end{tabular}
\label{table2}
\end{equation}
Again, the correspondence is one-to-one. 

In the analysis of the terms in the
OPE some simplification comes from the parity 
symmetry under exchange of $F^{+}$
and $F^{-}$ (and $\varepsilon _{\mu \nu \rho \sigma }\rightarrow
-\varepsilon _{\mu \nu \rho \sigma }$). Moreover, one has to use
systematically the special identity appearing in the second line of the
table and similar identities obtained by differentiating it. This
relationship follows by the properties of the $\varepsilon $-tensor (in
particular that the product of two $\varepsilon $-tensor is a quartic
polynomial is $\delta _{\mu \nu }$). Moreover, the existence of this
relationship is guaranteed by the following argument. $F^{+}$ and $F^{-}$
contain 3 independent components each. Therefore the bilinear $F_{\mu \nu
}^{+}F_{\rho \sigma }^{-}+F_{\rho \sigma }^{+}F_{\mu \nu }^{-}$ contains
nine independent components. Then it is necessarily re-expressible
algebraically via the stress-tensor, since the stress tensor contains
precisely nine independent components and it is bilinear in $F^{+}$-$F^{-}$.

To read the odd-spin content one has to use replacements of the type
\[
F_{\mu \nu }^{+}\overleftrightarrow{\partial _{\alpha }}F_{\rho \sigma
}^{-}-F_{\rho \sigma }^{+}\overleftrightarrow{\partial _{\alpha }}F_{\mu \nu
}^{-}\rightarrow -\frac{1}{2}\left( \varepsilon _{\mu \nu \rho \beta }{\cal A%
}_{\alpha \beta \sigma }^{(3)}-\varepsilon _{\mu \nu \sigma \beta }{\cal A}%
_{\alpha \beta \rho }^{(3)}-\varepsilon _{\rho \sigma \mu \beta }{\cal A}%
_{\alpha \beta \nu }^{(3)}+\varepsilon _{\rho \sigma \nu \beta }{\cal A}%
_{\alpha \beta \mu }^{(3)}\right) .
\]

The projector $\prod^{(2,2;4)}$ is the unique linear combination of the
three invariants (\ref{the3}) that is a polynomial of degree at most four in
uncontracted derivatives. This is easily seen by counting the free indices
and by the structure of the vector propagator. Therefore we have for a
vector 
\begin{eqnarray*}
{\prod }_{\mu \nu ,\rho \sigma ;\alpha \beta \gamma \delta }^{(2,2;4)} &=&%
\frac{1}{2^{9}\cdot 3\cdot 5}\left[ {\prod }_{\mu \nu ,\rho \sigma
}^{(2)}\left( \partial _{\alpha }\partial _{\beta }\partial _{\gamma
}\partial _{\delta }-{\rm tr}_{(\alpha \beta \gamma \delta )}\right) \right. 
\\
&&\left. -\frac{25}{14}\sum\Sb {\rm symm} \\ (\alpha \beta \gamma \delta )
\endSb {\prod }_{\mu \nu \alpha ,\rho \sigma \beta }^{(3)}\left( \partial
_{\gamma }\partial _{\delta }-\frac{1}{4}\Box \delta _{\gamma \delta
}\right) +\frac{5}{24}{\prod }_{\mu \nu \rho \sigma ,\alpha \beta \gamma
\delta }^{(4)}\right] .
\end{eqnarray*}
Let us now discuss the new terms, i.e. those with odd spin. The unique
spin-1 invariant is axial and has the form 
\begin{equation}
{\prod }_{\mu \nu ,\rho \sigma ;\alpha }^{(2,2;1)}=\sum_{{\rm symm}%
}\varepsilon _{\mu \rho \alpha \delta }~\pi _{\nu \sigma }~\partial _{\delta
}.  \label{spinor1}
\end{equation}
The symmetrization acts also on $\mu,\nu$ and $\rho,\sigma$.
However, there is no spin-1 axial current for the vector field. For the
spin-3 invariant ${\prod }_{\mu \nu ,\rho \sigma ;\alpha \beta \gamma
}^{(2,2;3)}$ there are two candidates:
\[
\sum_{{\rm symm}}\varepsilon _{\mu \rho \alpha \delta }~\pi _{\nu \sigma
}~\partial _{\beta }\partial _{\gamma }\partial _{\delta },\qquad \sum_{{\rm %
symm}}\varepsilon _{\mu \rho \alpha \delta }~\pi _{\nu \beta }~\pi _{\sigma
\gamma }~\partial _{\delta }.
\]
The combination appearing in (\ref{spi}) is the one with the least numer of
uncontracted derivatives. Similarly, the spin-5 invariant ${\prod }_{\mu \nu
,\rho \sigma ;\alpha \beta \gamma \varepsilon \delta }^{(2,2;5)}$ admits two
candidates, obtained by acting with two derivatives on the spin-3 ones:
\[
\sum_{{\rm symm}}\varepsilon _{\mu \rho \alpha \zeta }~\pi _{\nu \sigma
}~\partial _{\beta }\partial _{\gamma }\partial _{\delta }\partial
_{\varepsilon }\partial _{\zeta },\qquad \sum_{{\rm symm}}\varepsilon _{\mu
\rho \alpha \zeta }~\pi _{\nu \beta }~\pi _{\sigma \gamma }~\partial
_{\delta }\partial _{\varepsilon }\partial _{\zeta }.
\]
Again, the combination in (\ref{spi}) has the minimum number of derivatives.
Three odd-spin invariants (spin-1, spin-3 and spin-5) are not used for the
vector field: they will appear in the $TT$ OPE for the spinor. Indeed, the
general rule is that {\it all} independent invariants that one can construct
correspond to a nontrivial OPE term.

The spin-3, 4 and 5 two-point functions are 
\begin{eqnarray}
<{\cal A}_{\mu \nu \rho }^{(3)}(x)~{\cal A}_{\alpha \beta \gamma }^{(3)}(0)>
&=&{\frac{1}{2^{2}\cdot 3\cdot 7}}\left( \frac{1}{4\pi ^{2}}\right) ^{2}{%
\prod }_{\mu \nu \rho ,\alpha \beta \gamma }^{(3)}\left( {\frac{1}{|x|^{4}}}%
\right) ,  \nonumber \\
<{\cal J}_{\mu \nu \rho \sigma }^{(4)}(x)~{\cal J}_{\alpha \beta \gamma
\delta }^{(4)}(0)> &=&{\frac{1}{2^{2}\cdot 3^{2}\cdot 7}}\left( \frac{1}{%
4\pi ^{2}}\right) ^{2}{\prod }_{\mu \nu \rho \sigma ,\alpha \beta \gamma
\delta }^{(4)}\left( {\frac{1}{|x|^{4}}}\right) , \\
<{\cal A}_{\mu \nu \rho \sigma \tau }^{(5)}(x)~{\cal A}_{\alpha \beta \gamma
\delta \varepsilon }^{(5)}(0)> &=&\frac{1}{2^{2}\cdot 3\cdot 5\cdot 11}%
\left( \frac{1}{4\pi ^{2}}\right) ^{2}{\prod }_{\mu \nu \rho \sigma \tau
,\alpha \beta \gamma \delta \varepsilon }^{(5)}\left( {\frac{1}{|x|^{4}}}%
\right) .  \nonumber
\end{eqnarray}

\subsection{Spinor.}

The case of the spinor is the most involved. The stress tensor

\[
T_{\mu \nu }=\frac{1}{4}\left( \overline{\psi }\gamma _{\mu }%
\overleftrightarrow{\partial }_{\nu }\psi +\overline{\psi }\gamma _{\nu }%
\overleftrightarrow{\partial }_{\mu }\psi \right) ,\quad <\psi (x)~\overline{%
\psi }(0)>=\frac{1}{2\pi ^{2}}\frac{x\!\!\!\slash }{|x|^{4}}, 
\]
generates the OPE

\begin{eqnarray}
T_{\mu \nu }(x)T_{\rho \sigma }(y) &=&\frac{1}{10}~\left( \frac{1}{4\pi ^{2}}%
\right) ^{2}~{\prod }_{\mu \nu ,\rho \sigma }^{(2)}\left( \frac{1}{|x-y|^{4}}%
\right)   \nonumber \\
&&-\frac{1}{4\pi ^{2}}~{\cal A}_{\alpha }^{(1)}~\frac{1}{2}\sum_{{\rm symm}%
}\varepsilon _{\mu \rho \alpha \beta }~\pi _{\nu \sigma }~\partial _{\beta
}~\left( \frac{1}{|x-y|^{2}}\right)   \nonumber \\
&&+\frac{1}{4\pi ^{2}}~T_{\alpha \beta }~{\prod }_{\mu \nu ,\rho \sigma
;\alpha \beta }^{(2,2;2)}\left( |x-y|^{2}\ln |x-y|^{2}M^{2}\right)  
\nonumber \\
&&-\frac{1}{4\pi ^{2}}~{\cal A}_{\alpha \beta \gamma }^{(3)}~\frac{1}{2^{6}}%
\sum_{{\rm symm}}\varepsilon _{\mu \rho \alpha \delta }~\pi _{\nu \beta
}~\pi _{\sigma \gamma }~\partial _{\delta }{~}\left( |x-y|^{2}\ln
|x-y|^{2}M^{2}\right)   \label{spi} \\
&&+\frac{1}{4\pi ^{2}}~{\cal J}_{\alpha \beta \gamma \delta }^{(4)}~{\prod }%
_{\mu \nu ,\rho \sigma ;\alpha \beta \gamma \delta }^{(2,2;4)}\left(
|x-y|^{6}\ln |x-y|^{2}M^{2}\right)   \nonumber \\
&&-\frac{1}{4\pi ^{2}}~{\cal A}_{\alpha \beta \gamma \delta \varepsilon
}^{(5)}~\frac{1}{2^{13}\cdot 3^{2}}\sum_{{\rm symm}}\varepsilon _{\mu \rho
\alpha \zeta }~\pi _{\nu \beta }~\pi _{\sigma \gamma }~\partial _{\delta
}\partial _{\varepsilon }\partial _{\zeta }\left( |x-y|^{6}\ln
|x-y|^{2}M^{2}\right)   \nonumber \\
&&+{\rm descendants}+{\rm regular\ terms.}  \nonumber
\end{eqnarray}
Let us describe the even-spin terms first.

The term containing the stress tensor is a polynomial of degree at most four
in uncontracted derivatives. This fixes the relative coefficient of the
second and third invariant in the sum

\begin{eqnarray}
&&{\prod }_{\mu \nu ,\rho \sigma ;\alpha \beta }^{(2,2;2)}\left(
|x-y|^{2}\ln |x-y|^{2}M ^{2}\right) ={\rm SP}_{\mu \nu ,\rho \sigma ;\alpha
\beta }\left( \frac{1}{|x-y|^{2}}\right)  \nonumber \\
&&+\frac{9}{64}~{\prod }_{\mu \nu ,\rho \sigma }^{(2)}\partial _{\alpha
}\partial _{\beta }\left( |x-y|^{2}\ln |x-y|^{2}M ^{2}\right)  \nonumber \\
&&-\frac{15}{64}~{\prod }_{\mu \nu \alpha ,\beta \rho \sigma }^{(3)}\left(
|x-y|^{2}\ln |x-y|^{2}M ^{2}\right) .\quad \quad  \label{fg}
\end{eqnarray}
Another check is the universality of the coefficient of SP.

The spin-4 term is explicitly computed to give 
\begin{eqnarray*}
{\prod }_{\mu \nu ,\rho \sigma ;\alpha \beta \gamma \delta }^{(2,2;4)} &=&-%
\frac{1}{2^{14}\cdot 3^{{}}\cdot 5}\left[ {\prod }_{\mu \nu ,\rho \sigma
}^{(2)}\left( \partial _{\alpha }\partial _{\beta }\partial _{\gamma
}\partial _{\delta }-{\rm tr}_{(\alpha \beta \gamma \delta )}\right) \right. 
\\
&&\left. -\frac{25}{7}\sum\Sb {\rm symm} \\ (\alpha \beta \gamma \delta )
\endSb {\prod }_{\mu \nu \alpha ,\rho \sigma \beta }^{(3)}\left( \partial
_{\gamma }\partial _{\delta }-\frac{1}{4}\Box \delta _{\gamma \delta
}\right) +\frac{10}{3}{\prod }_{\mu \nu \rho \sigma ,\alpha \beta \gamma
\delta }^{(4)}\right] .
\end{eqnarray*}

The odd-spin terms are identified by the projector-invariants classified in
the previous subsection. Here there is no restriction on the number of
uncontracted derivatives and the spin-1 current is also present. The results
confirm the prediction that there is some OPE\ term for all existing
projector-invariants with the given properties.

The derivation of the OPE for the fermion, although more complicated than
the previous ones, is done following the same strategy. The higher spin
operators appearing on the right hand side form a basis for the monomials $%
\partial ^{p}\overline{\psi }\gamma _{\mu }\partial ^{q}\psi ,$ $p+q=$odd,
and $\partial ^{p}\overline{\psi }\gamma _{5}\gamma _{\mu }\partial ^{q}\psi
,$ $p+q=$even. Here proper symmetrizations in the indices are understood and
all derivatives are meant uncontracted, both among themselves and with the
index $\mu $ of the Dirac matrix. The reason for the restrictions on the
parity of $p+q$ is that the operators that do not satisfy it, like for
example the vector current $\overline{\psi }\gamma _{\mu }\psi $ ($p+q=0=$%
even) do not appear in the operator product expansion. This is clearly seen
by decomposing the Dirac fermion $\psi =\psi _{1}+i\psi _{2}$ into its
Majorana components $\psi _{1,2}$ and observing that the stress tensor and
the propagator do not mix the two. In conclusion, a one-to-one
correspondence table like (\ref{table1}) and (\ref{table2}) can be worked
out. We do not write the complete table. It is sufficient to report the
primary spin-$\left( p+q+1\right) $ contents of each monomial $\partial
^{p}(\gamma _{5})\overline{\psi }\gamma _{\mu }\partial ^{q}\psi $. We
observe that $\partial ^{p}\overline{\psi }(\gamma _{5})\gamma _{\mu
}\partial ^{q}\psi $ is equivalent to $-\partial ^{p-1}\overline{\psi }%
(\gamma _{5})\gamma _{\mu }\partial ^{q+1}\psi $ for our purposes, since the
difference of the two is a descendant of a spin-$\left( p+q\right) $
operator. Up to descendants and permutations of the indices, we have, very
simply, 
\[
\begin{tabular}{|c|c|c|c|c|}
\hline
$\overline{\psi }\gamma _{5}\gamma _{\mu }\psi $ & $\overline{\psi }\gamma
_{\mu }\partial _{\nu }\psi $ & $\overline{\psi }\gamma _{5}\gamma _{\mu
}\partial _{\nu }\partial _{\rho }\psi $ & $\overline{\psi }\gamma _{\mu
}\partial _{\nu }\partial _{\rho }\partial _{\sigma }\psi $ & $\overline{%
\psi }\gamma _{5}\gamma _{\mu }\partial _{\nu }\partial _{\rho }\partial
_{\sigma }\partial _{\alpha }\psi $ \\ \hline
${\cal A}_{\mu }^{(1)}$ & $T_{\mu \nu }$ & $\frac{1}{4}{\cal A}_{\mu \nu
\rho }^{(3)}$ & $\frac{1}{8}{\cal A}_{\mu \nu \rho \sigma }^{(4)}$ & $\frac{1%
}{16}{\cal A}_{\mu \nu \rho \sigma \alpha }^{(5)}$ \\ \hline
\end{tabular}
\]

We have computed the two-point functions. The results are 
\begin{eqnarray*}
<{\cal A}_{\mu }^{(1)}(x)~{\cal A}_{\alpha }^{(1)}(0)> &=&\frac{2^{2}}{3}%
\left( \frac{1}{4\pi ^{2}}\right) ^{2}{\pi }_{\mu \alpha }\left( {\frac{1}{%
|x|^{4}}}\right) , \\
<{\cal A}_{\mu \nu \rho }^{(3)}(x)~{\cal A}_{\alpha \beta \gamma }^{(3)}(0)>
&=&\frac{2^{4}}{3\cdot 5\cdot 7}\left( \frac{1}{4\pi ^{2}}\right) ^{2}{\prod 
}_{\mu \nu \rho ,\alpha \beta \gamma }^{(3)}\left( {\frac{1}{|x|^{4}}}%
\right) , \\
<{\cal J}_{\mu \nu \rho \sigma }^{(4)}(x)~{\cal J}_{\alpha \beta \gamma
\delta }^{(4)}(0)> &=&\frac{2^{2}}{3^{2}\cdot 7}\left( \frac{1}{4\pi ^{2}}%
\right) ^{2}{\prod }_{\mu \nu \rho \sigma ,\alpha \beta \gamma \delta
}^{(4)}\left( {\frac{1}{|x|^{4}}}\right) , \\
<{\cal A}_{\mu \nu \rho \sigma \tau }^{(5)}(x)~{\cal A}_{\alpha \beta \gamma
\delta \varepsilon }^{(5)}(0)> &=&\frac{2^{5}}{3\cdot 5\cdot 7\cdot 11}%
\left( \frac{1}{4\pi ^{2}}\right) ^{2}{\prod }_{\mu \nu \rho \sigma \tau
,\alpha \beta \gamma \delta \varepsilon }^{(5)}\left( {\frac{1}{|x|^{4}}}%
\right) .
\end{eqnarray*}
The number theoretical relationship between the spin of a current and the
prime factors is natural: no prime number grater than $2s+1$ appears for
spin $s$ (the one-loop values of Feynman diagrams are computable
algebraically). Finally, observe that our two-point functions 
are reflection positive.

\subsection{Classical and quantum conformal algebras.}

All the invariants appearing in the OPE, but SP, are annihilated {\it %
algebraically} by $\partial _{\mu }$ or by the trace contraction $\delta
_{\mu \nu }$. Instead, SP is not annihilated algebraically and a contact
term survives. Explicitly, 
\[
\partial _{\mu }{\rm SP}_{\mu \nu ,\rho \sigma ;\alpha \beta }={\frac{1}{2}}%
[-2\delta _{\alpha \rho }\delta _{\beta \sigma }\partial _{\nu }+\partial
_{\alpha }(\delta _{\beta \rho }\delta _{\nu \sigma }+\delta _{\beta \sigma
}\delta _{\nu \rho }-\delta _{\beta \nu }\delta _{\rho \sigma })]\Box
+(\alpha \leftrightarrow \beta )-{\rm tr}_{\alpha \beta }. 
\]
The surviving local terms describe the symmetries of the {\it classical}
conformal algebra. Let us show this explicitly.

Keeping the contact terms originated by $\Box (1/|x|^{2})$ we obtain 
\begin{equation}
T(x)~T_{\rho \sigma }(y)=2\delta (x-y)~T_{\rho \sigma },  \label{teta}
\end{equation}
and 
\begin{eqnarray*}
\partial _{\mu }T_{\mu \nu }(x)~T_{\rho \sigma }(y) &=&-\partial _{\alpha
}\delta (x-y)~\left[ \delta _{\nu \sigma }T_{\alpha \rho }+\delta _{\nu \rho
}T_{\alpha \sigma }-\delta _{\rho \sigma }T_{\alpha \nu }\right] \\
&&+2\partial _{\nu }\delta (x-y)~T_{\rho \sigma }+1^{st}~{\rm descendant.}
\end{eqnarray*}
We have to study the first descendant, since $\partial T$ can multiply $%
\delta (x-y)$. Further descendants can be neglected, because they cannot
multiply local terms (accordingly, the transformation rule does cannot
contain derivatives of $T$ beyond the first). One can check that the first
descendant can be set to zero if $T$ is located in $(x+y)/2.$

We have $T_{\mu \nu }={\frac{2}{\sqrt{g}}}{\frac{\delta {\cal S}}{\delta
g^{\mu \nu }}}$, where ${\cal S}$ denotes the action. In the linear
approximation, ${\cal S}={\cal S}_{0}-{\frac{1}{2}}\phi _{\mu \nu }T_{\mu
\nu }$, where $g_{\mu \nu }=\delta _{\mu \nu }+\phi _{\mu \nu }$.
Dilatations correspond to the variation $\delta _{\Lambda }\phi _{\mu \nu
}=2\Lambda \delta _{\mu \nu }$. Consequently, acting on the exponential $%
{\rm e}^{-{\cal S}}$ in the functional integral, the operator $\delta
_{\Lambda }$ inserts $\int d^{4}x~\Lambda T$ in the correlators. Therefore, $%
\delta _{\Lambda }T_{\mu \nu }=2\Lambda T_{\mu \nu },$ from (\ref{teta}).

Diffeomorphisms are expressed by $\delta _{\xi }\phi _{\mu \nu }=\partial
_{\mu }\xi _{\nu }+\partial _{\nu }\xi _{\mu }$: the operator $\delta _{\xi
} $ inserts $\int d^{4}x~(\partial _{\mu }\xi _{\nu }~T_{\mu \nu })$ in the
correlators. Therefore, 
\begin{eqnarray}
\delta _{\xi }T_{\rho \sigma }(y) &=&\int d^{4}x~\partial _{\mu }\xi _{\nu
}(x)~T_{\mu \nu }(x)~T_{\rho \sigma }(y)=-\int d^{4}x~\xi _{\nu
}(x)~\partial _{\mu }T_{\mu \nu }(x)~T_{\rho \sigma }(y)  \nonumber \\
&=&\xi ^{\alpha }~\partial _{\alpha }T_{\rho \sigma }-\partial _{\alpha }\xi
_{\rho }~T_{\alpha \sigma }-\partial _{\alpha }\xi _{\sigma }~T_{\rho \alpha
}+2~\partial \cdot \xi ~T_{\rho \sigma }+\delta _{\rho \sigma }~\partial
_{\alpha }\xi _{\beta }~T_{\alpha \beta },  \label{dif}
\end{eqnarray}
after integrating by parts. The transformation rule is not the expected one.
The reason is that there can be local terms in the OPE other than those
obtained by the above procedure. i.e. SP is defined up to the local terms: 
\begin{eqnarray*}
T_{\mu \nu }(x)~T_{\rho \sigma }(y) &\rightarrow &T_{\mu \nu }(x)~T_{\rho
\sigma }(y)+A~\delta (x-y)\left( \delta _{\mu \nu }T_{\rho \sigma }+\delta
_{\rho \sigma }T_{\mu \nu }\right) \\
&&+B~\delta (x-y)\left( \delta _{\mu \rho }T_{\nu \sigma }+\delta _{\mu
\sigma }T_{\rho \nu }+\delta _{\nu \rho }T_{\mu \sigma }+\delta _{\nu \sigma
}T_{\mu \rho }\right) .
\end{eqnarray*}
(\ref{teta}) was correct and in order to preserve it one has to impose $%
A+B=0 $. Choosing $B=1$ in order to cancel the last term of (\ref{dif}), we
obtain 
\[
\delta _{\xi }T_{\rho \sigma }(y)=\xi ^{\alpha }~\partial _{\alpha }T_{\rho
\sigma }+\partial _{\rho }\xi ^{\alpha }~T_{\alpha \sigma }+\partial
_{\sigma }\xi ^{\alpha }~T_{\rho \alpha }+~\partial \cdot \xi ~T_{\rho
\sigma }, 
\]

The quantities that appear in the OPE (central charge, operators of various
spin, as well as descendants and regular terms) constitute the {\it quantum}
conformal algebra \cite{noialtri}. In two dimensions the difference between
quantum and classical conformal algebras is just the identity operator (and
therefore there is a unique, primary, central charge). In general dimension
the difference is a tower of operators, classified by their spin (and
another label $I$), defining infinitely many (secondary) central charges. In
the next subsection we write down the general quantum conformal algebra for
interacting conformal field theories and discuss its properties.

\subsection{Interacting critical theories.}

The free-field OPE\ can be easily generalized to give the structure of the $%
TT$ OPE in the most general conformal field theory. We recall that the
projectors ${\prod }_{\mu \nu ,\rho \sigma ;\alpha _{1}\cdots \alpha
_{s}}^{(2,2;s)}$ do not satisfy conservation in the indices $\alpha
_{1}\cdots \alpha _{s}$ and this means that the operators appearing on the
right hand side of the OPE\ can indeed be non-conserved. For the reasons
that we have explained in sect. 2, the only violation of conservation
allowed at criticality is the explicit $h$-violation due to the anomalous
dimension $h$ of the operator in question. Indeed, the structure of the OPE
remains unchanged in presence of an anomalous dimension $h$, since the
projectors that we have classified can act on any power of $|x-y|$.
Actually, this is true for all of them but the special one, which has
necessarily to act on $1/|x|^{2}$ in order to produce the desired $\delta $%
-functions. This statement is precisely the finiteness of the stress tensor,
implied by the Poincar\'{e} algebra. Therefore the special term (and its
descendants) is fixed, while all other terms depend on the theory. There are
as many independent operators as projectors ${\prod }^{(2,2;s)},$ at least.

The operators associated with the projectors ${\prod }^{(2,2;s)}$ are
primary and correlators of primary operators are conformal. Traces, which
are indeed primary operators of lower spin (and the same dimension), are
appropriately extracted from ${\prod }^{(2,2;s)}$. Improvement terms \'{a}
la (\ref{impro}) are descendants of lower spin currents and therefore cannot
appear multiplied by ${\prod }^{(2,2;s)}$ (this would violate conformal
symmetry, see (\ref{c2})). They can appear only multiplied by projector
descendants \`{a} la (\ref{desce}). Projector-primaries identify
primary-operators and vice versa. This solves the problem of identifying the
appropriate currents at criticality, see the end of section 3.

We have seen that for a given $s$ the projector ${\prod }^{(2,2;s)}$ is not
unique, actually in general there is a triple degeneracy in the free-field
limit. Actually, this triple degeneracy can be enhanced in presence of
internal symmetries, when there are matter fields in various different
irreducible representations of the gauge group, etc. Therefore we have to
introduce a second label $I$ and sum over it. Correspondingly, there will be
central charges $c_{s}^{I}$ for each channel and each spin (see next
section). The resulting structure is

\begin{eqnarray}
T_{\mu \nu }(x)~T_{\rho \sigma }(y) &=&\frac{1}{60}~\left( \frac{1}{4\pi ^{2}%
}\right) ^{2}~{\prod }_{\mu \nu ,\rho \sigma }^{(2)}\left( \frac{c_{2}}{%
|x-y|^{4}}\right) +{\frac{1}{4\pi ^{2}}}~T_{\alpha \beta }~{\rm \tilde{SP}}_{\mu \nu
,\rho \sigma ;\alpha \beta }\left( x-y\right)  \nonumber \\
&&\!\!\!\!\!\!\!\!\!\!\!\!\!\!\!\!\!\!\!\!\!\!\!\!\!\!\!\!\!\!\!\!\!\!\!\!\!%
\!\!\!\!\!{+~{\frac{1}{4\pi ^{2}}}\sum\Sb s,I:  \\ s-6+h_{s,I}<0  \endSb 
{\cal J}_{\alpha _{1}\cdots \alpha _{s}}^{(s,I)}~~{\prod }_{\mu \nu ,\rho
\sigma ;\alpha _{1}\cdots \alpha _{s}}^{(2,2;s,I)}\left[
|x-y|^{2s-2+h_{s,I}}\mu ^{h_{s,I}}\left( \ln |x-y|^{2}M^{2}\right) ^{\sigma
(2s-2+h_{s,I})}\right] }  \nonumber \\
&+&{\rm descendants}~+~{\rm regular\ terms.}  \label{hope}
\end{eqnarray}
This section is devoted to the analysis of this formula.

The factor $\mu ^{h_{s,I}}$ is needed to match the dimensions and assures
the compatibility of the renormalization group equations ${\rm d}(TT)/\,{\rm %
d}\ln \mu =0$ and ${\rm d}{\cal J}^{(s,I)}/\,{\rm d}\ln \mu =-h_{s,I}\,{\cal %
J}^{(s,I)}$.

The label $I$ runs over the projectors of formula (\ref{thetwo}) for $s=2$
and over those of formula (\ref{the3}) for $s=4$, as well as over eventual
internal indices. Kinematically, projectors (\ref{thetwo}) and (\ref{the3})
are sufficient to emphasize the presence of the $I$-degeneracy, but only
dynamics can determine it precisely. Let us explain this point in detail.

Operators with different spin are orthogonal, according to property (\ref
{ics}). However, the situation is more complicated for operators with the
same spin. There is no unique definition of the projectors ${\prod }_{\mu
\nu ,\rho \sigma ;\alpha _{1}\cdots \alpha _{s}}^{(2,2;s,I)}$:\ any linear
combination of the projectors of (\ref{thetwo}) and (\ref{the3}) is allowed.
Correspondingly, the true operators ${\cal J}_{\alpha _{1}\cdots \alpha
_{s}}^{(s,I)}$ are linear combinations of the ${\cal J}_{\alpha _{1}\cdots
\alpha _{s}}^{(s)}$-operators of scalar, spinor and vector fields. Only
dynamics can determine the precise nature of the labels $I,$ and so remove
this degeneracy of the free-field limit.

In general, orthogonal combinations ${\cal J}_{\alpha _{1}\cdots \alpha
_{s}}^{(s,I)}$ with the same $s$ will be distinguished by their different
anomalous dimensions $h_{s,I}$. Operators with different dimensions are
clearly orthogonal and define different channels $I$, but in interacting CFT$%
_{4}$'s there might be operators with the same spin and the same dimension.
In that case it is more convenient to triangulate the matrix of two-point
functions $M_{s}^{II^{\prime }}=<{\cal J}^{(s,I)}{\cal J}^{(s,I^{\prime })}>$
rather than diagonalize it. Finally, there is the case of operators with the
same spin, the same dimension and the same channel (i.e. the same $I$).
There is now way to remove this residual degeneracy, which is the mixing
studied in sect. 2 of \cite{ccfis}: the set of mixing operators ${\cal O}%
_{i} $ with these properties define a unique central charge. Issues like
this will be debated at length in the next section.

The tilde on the special projector-invariant SP denotes that SP itself is
identified up to linear combinations of the other two
projector-invariants  ${\prod }^{(2,2;2)}$. The combination
depends on the theory.

From the general form of the OPE\ we observe that the higher spin symmetry
is precisely the infinite symmetry characterizing free field theory, since
only when all $h$'s are zero the higher spin currents are conserved. The set
of operators appearing in the OPE is different for scalar, spinor and vector
fields, so we can say that there are three known types of such infinite
symmetries. There could be more, in particular in the domain of higher
derivative theories \cite{antonia}, where it might be possible to preserve
some flavor higher spin symmetry in presence of certain interactions. This
research is under consideration.

In conclusion, we identify the primary currents of spin $s$ via the $TT$
OPE. This we can do, for the moment, in the UV\ and IR\ limits, i.e. we know
how to study the initial and final ``positions'' CFT$_{UV}$ and CFT$_{IR}$
of the RG trajectory. In the next section we will discuss the off-critical
extension of this analysis. The reader should keep in mind the general
viewpoint described in the introduction, since we are proceeding precisely
as represented in the second part of Fig. 1.

We now make several comments on interacting conformal field theories.
Concrete examples in which one can study these issues using perturbation
theory are provided by supersymmetric theories. The most singular term in
the $TT$ OPE after the central charge contains the lowest component $%
\overline{\varphi }\varphi =\Sigma $ of the Konishi superfield \cite
{noialtri}, which is indeed anomalous dimensioned, 
\[
TT\sim \frac{c}{|x|^{8}}+\frac{\overline{\varphi }\varphi }{|x|^{6-h}}%
+\cdots 
\]
This is true in any N=1 supersymmetric theory (so, also off-criticality; see 
\cite{c'} for the geeral expression of $h$ to the lowest order). Superfield
formulas are presented in the appendix and they relate the free field
currents of scalar, spinor and vector fields. Other components of the
Konishi N=1 superfield are the Konishi current (which is basically ${\cal A}%
^{(1)}$ in the notation of the present paper) and the Kinetic Lagrangian for
matter multiplets. Therefore these operators are not conserved and have the
same anomalous dimension $h$.

Now, let us consider N=4 supersymmetric Yang-Mills theory. The extended
supersymmetry relates $\overline{\varphi }\varphi $ to various other
operators (the components of a hypothetical ``N=4 superfield'') not
contained in the N=1 Konishi superfield \cite{fer}. Among these operators
there is one of the three spin-4 currents (already reduced to two by
supersymmetry) that appear in the OPE, as well as a spin-3 current and other
components.

Therefore, the ``secondary central charge'' $c^{\prime }$ studied in \cite
{c'} at the second loop order in perturbation theory is also one of the
spin-4 central charges in the language of the present paper. Moreover, the
anomalous dimension $h$ computed in \cite{noialtri} and \cite{c'} for the
Konishi operator is also the anomalous dimension of the spin-4 current \cite
{fer}. Since $h$ is nonvanishing already at the one loop level, precisely $%
h=3N_c\alpha/\pi$ \cite{noialtri}, this clearly shows that the spin-4
current is not conserved even if the theory is conformal. The Konishi N=4
supermultiplet does not contain a spin-5 current \cite{fer}, so there is
necessarily a further superfield appearing in the singular terms of the OPE.
Actually, one can show that there are two \cite{new}.

\vskip .2truecm 

%%Begin InstantTeX Picture
\let\picnaturalsize=N

%If you do not have the picture file add:
%\let\nopictures=Y
%to the beginning of the file.
\ifx\nopictures Y\else{\ifx\epsfloaded Y\else\fi
\global\let\epsfloaded=Y \centerline{\ifx\picnaturalsize N\epsfxsize
6.0in\fi \epsfbox{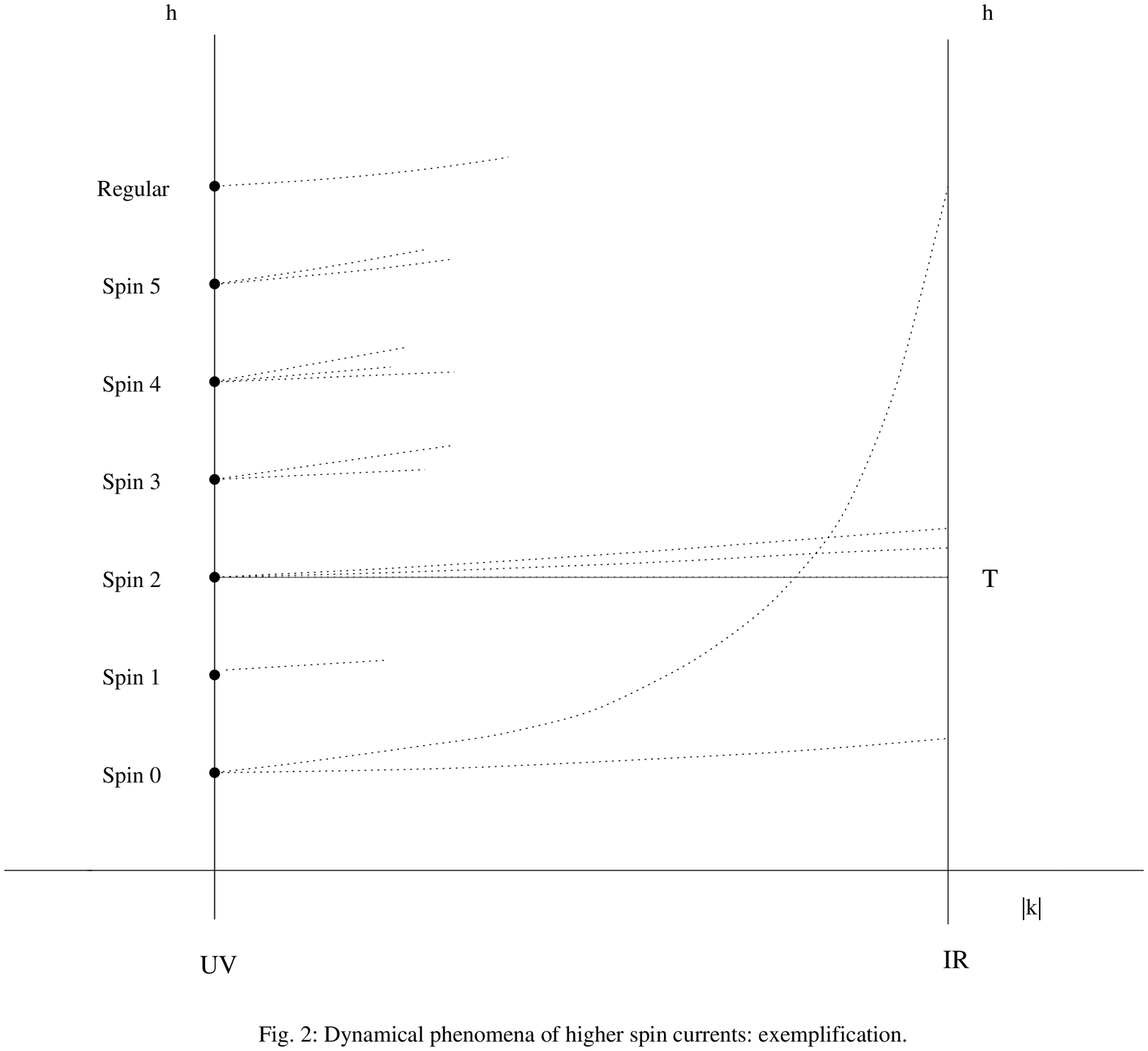}}}\fi
%%End InstantTeX Picture
\vskip .3truecm

In the strongly coupled
large $N_{c}$ limit, according to a recent conjecture \cite{malda},
there is a certain correspondence between N=4 supersymmetric Yang-Mills
theory with gauge group $SU(N_{c})$ and supergravity on Anti de Sitter
space. Calculations based on this correspondence \cite{kleb} show that $h$
should grow indefinitely in this limit and therefore that the mixing terms
should disappear from the OPE \cite{fer}. Assuming that this is correct, we
see that the OPE (\ref{hope}) with $h_{s,I}=\infty $ will contain only the
identity operator and the stress tensor itself (the special term SP),
together with its descendants. We conclude that conformal field theories in
the strongly coupled 
large $N_{c}$ limit are the best analogues of two dimensional conformal
field theories. Correspondingly, they should be much more symmetric than
ordinary conformal field theories in four dimensions. It is compulsory to
look for an eventual ``infinite symmetry'' characterizing these special
theories \cite{pala}, presumably rather different from the higher spin
symmetries of free field theories. Higher spin flavour symmetry
could be the key-concept for a complete classification of conformal field theories
in four dimensions.

We have stressed that an interesting phenomenon occurring in our
investigation is the splitting of the $I$-degeneracy, depicted in Fig. 2. We
do not have examples in families of conformal field theories. Presumably it
is possible to construct such examples in finite N=2 theories. We do have
examples for RG\ flows interpolating between pairs of CFT$_{4}$'s. In
particular, there is one case in which this phenomenon is evident \cite
{noialtri,andrei}, also at the nonperturbative level. We briefly describe it.

Consider N=1 supersymmetric QCD with $N_{c}$ colors and $N_{f}$ flavors. In
the conformal window $3/2~N_{c}<N_{f}<3N_{c}$ this theory (called
``electric'') has an IR fixed point, expected to be equivalent to the IR\
fixed point of a very different theory, the ``magnetic'' dual \cite{seiberg}%
. The magnetic theory contains gluons, quarks and mesons, and a
superpotential. In the neighborhood $N_{f}\lesssim 3N_{c},$ the IR\ fixed
point is weakly coupled from the electric point of view and strongly coupled
from the magnetic point of view. From the electric point of view, the $TT$
OPE outlines a single operator $\Sigma ,$ with a nonvanishing small
anomalous dimension $h$. One can indeed use perturbation theory and the
unicity of $\Sigma $ follows from power counting. See \cite{noialtri} for
the value of $h$. Instead, the magnetic theory exhibits a case of $I$%
-degeneracy. In the UV\ limit the $TT$ OPE contains a single $\Sigma $
operator, which is however the sum of two Konishi currents, of the quarks
and of the mesons. These two Konishi currents have no reason to be parts of
a single operator. Perturbative calculations around the magnetic UV fixed
point show that indeed the two components of $\Sigma $ are splitted by
acquiring different (and both nonvanishing) anomalous dimensions.

The RG flow of the magnetic theory to the IR limit is beyond perturbation
theory. It can be studied by assuming electric-magnetic duality, which
implies that one (and only one) of the anomalous dimensions remains small,
while the other one, say $h_{m},$ grows enough to move the mesonic Konishi
operator $\Sigma _{m}$ very far away, actually in the regular terms, as we
now prove. This does not mean that $\Sigma _{m}$ has no electric companion,
rather that the electric partner is a higher dimensioned composite operator
(according to the identification $M_{j}^{i}=Q^{i}\tilde{Q}_{j}$ of the meson
field $M_{j}^{i}$ as a product of electric quark fields). In the limit $%
N_{f}=3N_{c},$ where the electric theory is free, $\Sigma _{m}$ disappears
from the electric OPE's (a free OPE does not contain it) and therefore $%
h_{m} $ tends to infinity in this limit. This phenomenon is similar to the
one occurring in the large $N_{c}$ limit of N=4 supersymmetric Yang-Mills
theory \cite{kleb}.

We have depicted this behavior in Fig. 2 (the two lines exiting from the dot
labelled ``spin 0''). Regular terms can become singular and other similar
situations can occur. The other situations appearing in Fig. 2 are purely
illustrative.

The operators appearing in the $TT$ OPE are natural deformations of the
theory \cite{noialtri}. This is particularly relevant in the supersymmetric
case where the anomalous dimension of the Konishi operator $\Sigma $ equals
the derivative of the beta-function at criticality. In the
non-supersymmetric case $\Sigma =\varphi ^{2}$ is just the mass deformation.
The other operators represent nontrivial axial-current deformations, spin-2
deformations, higher spin deformations. We finally remark that the
equivalence of two conformal field theories necessarily implies equivalence
of the quantum field theories obtained by these deformations (this is the
idea of e-m universality, see ref. \cite{ccfis}: the observation that a
critical theory carries nontrivial information about the quantum field
theories of which it can be the limit).

\section{Higher spin central charges and central functions.}

The definition of central functions $c_{s,I}(g)$ interpolating between the
critical values of the higher spin central charges proceeds as in sect. 2 of
ref. \cite{ccfis}. One identifies the higher spin current ${\cal J}^{(s,I)}$
as a channel of the stress-tensor four-point function $%
<T_{1}T_{2}T_{3}T_{4}> $ in the limit in which $T_{1}$ is close to $T_{2}$
and $T_{3}$ is close to $T_{4}$, see Fig. 3. The general form of the
two-point function $<{\cal J}^{(s,I)}{\cal J}^{(s,I)}>$, formula (\ref{auto2}%
), has to be combined with the $TT$ OPE studied in the previous section. We
do not repeat here the cut-and-paste construction of $c_{s,I}(g)$, but we
assume and use it. We just recall that this construction assures that $%
c_{s,I}(g)$ depends only on the running coupling constant $g$
notwithstanding the $h$-violation of the conservation condition, so that the
problem of sect. 3 is circumvented. We recall that the discussion of most of
that section cannot be applied to ordinary quantum field theory as it
stands, since it assumes the existence of a higher spin flavor symmetry.

Identifying ${\cal J}^{(s,I)}$ as a channel of correlators of conserved
currents eliminates a certain amount of arbitrariness in the definitions of $%
{\cal J}^{(s,I)}$ itself. For example, the overall constants of the
quantities $c_{s,I}(g)$ are fixed, since the external legs of our
correlators are always stress-tensors. Secondly, we had no way to fix the
relative factors in front of the scalar, spinor and vector contributions to
the currents. Now we know that this problem is related to the $I$-degeneracy
and that it is removed by stating that the appropriate combinations of
projectors and operators have to be read via the $TT$ OPE. As we said, the
dynamics of the theory is deeply involved, so one needs to go at least to
two loops to appreciate this effect. Thirdly, other nuisances, like scheme
dependence, are automatically bypassed by this construction and the
functions $c_{s,I}(g)$ are physical.

In practice, to make the comparison between
our construction and the notions commonly considered
in the theory of deep
inelastic scattering \cite{muta} we are replacing the electron and parton
with two or more gravitons, in order
to turn those techniques into useful tools for a theoretical
investigation of conformal windows and conformal field theories.
The central functions of \cite{ccfis} have therefore this
``physical'' interpretation. 

We consider a theory with $N_{0}$ real scalars, $N_{1/2}$ Dirac spinors and $%
N_{1}$ vector fields. The $\Sigma $-channel was studied extensively in \cite
{c'}, where the central charge $c_{0}$ was called $c^{\prime }:$ 
\[
c_{0}=N_{0}. 
\]
Precisely, le structure of the channel is 
\[
\frac{1}{18}~\left( \frac{1}{4\pi ^{2}}\right) ^{4}~{\prod }_{\mu \nu ,\rho
\sigma }^{(2)}\left( \frac{1}{|x-y|^{4}}\right) \frac{N_{0}}{\left|
(x+y-z-w)/2\right| ^{4}}~{\prod }_{\alpha \beta ,\gamma \delta }^{(2)}\left( 
\frac{1}{|z-w|^{4}}\right) . 
\]
In the spin-2 channel we observe the first case if $I$-degeneracy. There are
three operators, the stress tensors $T_{0},$ $T_{1/2}$ and $T_{1}$ of
scalar, spinor and vector fields. One combination is precisely the
stress-tensor of the theory, $T=$ $T_{0}+T_{1/2}+T_{1}$, identified by the
projector SP, with canonical dimension. $T$ does not mix with the other two
for obvious reasons and so we know a priori how this kind of $I$-degeneracy
is removed.

The central function $c_{(2)}^{T}$ associated with the $T$-channel of the
four-point function of the stress-tensor is just the function $c(g)$ of (\ref
{two}). The one-loop value is

\[
c_{2}^{T}=N_{0}+6N_{1/2}+12N_{1} 
\]
and the structure of the channel can be read by combining the appropriate
OPE term with the two-point function itself, 
\[
\frac{1}{60}\left( \frac{1}{4\pi ^{2}}\right) ^{4}~{\rm SP}_{\mu \nu ,\rho
\sigma ;\varepsilon \zeta }(x-y)~~{\prod }_{\varepsilon \zeta ,\iota \kappa
}^{(2)}\left( \frac{c_{2}^{T}}{\left| (x+y-z-w)/2\right| ^{4}}\right) ~{\rm %
SP}_{\alpha \beta ,\gamma \delta ;\iota \kappa }(z-w). 
\]

\setcounter{figure}{2} 
\begin{figure}[tbp]
\begin{center}
\begin{picture}(220,100)(0,10)
\thicklines

\put(45,20){\circle*{4}}
\put(45,70){\circle*{4}}
\put(175,20){\circle*{4}}
\put(175,70){\circle*{4}}

%\put(20,20){\line(0,1){50}}
%\put(200,20){\line(0,1){50}}

\put(70,45){\circle*{4}}
\put(150,45){\circle*{4}}

\put(45,20){\line(1,1){25}}
\put(45,70){\line(1,-1){25}}
\put(175,20){\line(-1,1){25}}
\put(175,70){\line(-1,-1){25}}

\put(70,45){\line(1,0){80}}

\put(110,55){\makebox(0,0){${\cal J}^{(4)}_{\alpha\beta\gamma\delta}$}}

\put(25,70){\makebox(0,0){$T_{\mu\nu}$}}
\put(25,20){\makebox(0,0){$T_{\rho\sigma}$}}
\put(195,70){\makebox(0,0){$T_{\alpha\beta}$}}
\put(195,20){\makebox(0,0){$T_{\gamma\delta}$}}
\end{picture}
\end{center}
\caption{Construction of $c_{4}$.}
\end{figure}
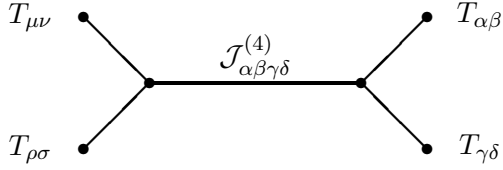
 
As we said, there are other two spin-2 central charges, corresponding to
combinations of the two invariants of (\ref{thetwo}). We cannot say which
precise combinations split the degeneracy since our analysis is one-loop,
but we can make some observations. We choose to present the one-loop
formulas for the operators ${\cal O}_{2}=T_{0}+T_{1/2}$ and ${\cal O}%
_{3}=T_{0}$ (we put ${\cal O}_{1}=T$) and we call the respective charges $%
c_{2}^{(0+1/2)}$ and $c_{2}^{(0)}$. The corresponding projectors, ${\prod }%
_{\mu \nu ,\rho \sigma ;\alpha \beta }^{(2,2;2)(sc+f)}$ and ${\prod }_{\mu
\nu ,\rho \sigma ;\alpha \beta }^{(2,2;2)(sc)}$ are, respectively, the sum
of the second and third term of formula (\ref{fg}) and the difference
between (\ref{ft}) and (\ref{fg}).

The purpose of this choice is to triangulate the matrix of two-point
functions, actually to impose an even further constraint $<{\cal O}_{i}{\cal %
O}_{j}>=<{\cal O}_{i}{\cal O}_{i}>$ for $i<j$. This constraint reduces the
set of independent entries to the diagonal ones. Instead, orthogonality $<%
{\cal O}_{i}{\cal O}_{j}>=0$ for $i\neq j$ would require awkward
denominators like $1/N_{0}$ and appears therefore meaningless (the simplest
orthogonal operators are just $T_{0},$ $T_{1/2}\ $and $T_{1}$ but dynamics
makes it compulsory to isolate $T$, due to its special nature; this shows
that the $I$-degeneracy is a nontrivial issue). We have simply 
\[
c_{2}^{(sc+f)}=N_{0}+6N_{1/2},\qquad c_{2}^{(sc)}=N_{0}. 
\]
The structure of the channel reads

\begin{eqnarray}
\frac{1}{60}\left( \frac{1}{4\pi ^{2}}\right) ^{4} &&{\prod }_{\mu \nu ,\rho
\sigma ;\varepsilon \zeta }^{(2,2;2,I)}\left( |x-y|^{2}\ln
|x-y|^{2}M^{2}\right) ~{\prod }_{\varepsilon \zeta ,\iota \kappa
}^{(2)}\left( \frac{c_{2}^{(I)}}{\left| (x+y-z-w)/2\right| ^{4}}\right) 
\nonumber \\
\times &&{\prod }_{\alpha \beta ,\gamma \delta ;\iota \kappa
}^{(2,2;2,I)}\left( |z-w|^{2}\ln |z-w|^{2}M^{2}\right) ,  \label{mid2}
\end{eqnarray}
with $I=sc+f$ or $I=sc$. We stress again that the issue of $I$-degeneracy
demands for a two-loop analysis.

The discussion can be repeated in the other cases. For odd spin there is no $%
I$-degeneracy at the level of our investigation, for spin-4 there is a
triple degeneracy. The formulas for the channels and charges can be written
straightforwardy following the general recipe.

Off-criticality the OPE contains several quantities and operators that
vanish or disappear at criticality. First of all, many more projectors
appear, since $T_{\mu \nu }$ is not traceless. These are new channels of the
many-point functions of the stress tensor. We can nevertheless focus on the
terms carried by our traceless projectors and treat the other ones apart
(i.e. in the OPE's $T_{\mu \nu }(x)T(y)$ and $T(x)T(y)$). More importantly,
there are new functions in the same channel, say ${\cal J}^{(s)}$. These are
the $c_{s,s^{\prime }}$ of section 3, formula (\ref{auto2}). They change the
internal structure of the channel (the middle factor of (\ref{mid2})), but
not the external structure. Combining the discussions of sections 3 and 4,
the full set of off-critical functions is labelled by the spin $s,$ the
mixing spin $s^{\prime }$ and the index $I.$ Precisely, $c_{s,s^{\prime
}}^{I},$ $s^{\prime }=s,\ldots 0,$ $c_{s,s}^{I}=c_{s}^{I}$. Each function
will have its own operator in the OPE, the associated operator component of
the ${\cal J}^{(s,I)}$-divergence (see also (\ref{deco})). For example, $%
c_{2,0}^{T}$ is the function called $f$ in formula (\ref{two}). Its operator
is the trace $\Theta $ of the stress-tensor (which indeed represents a
spin-2-0 mixing). $c_{2,0}^{T}$ is the central function associated with the $%
\Theta $-channel of the four-point function of the stress tensor.

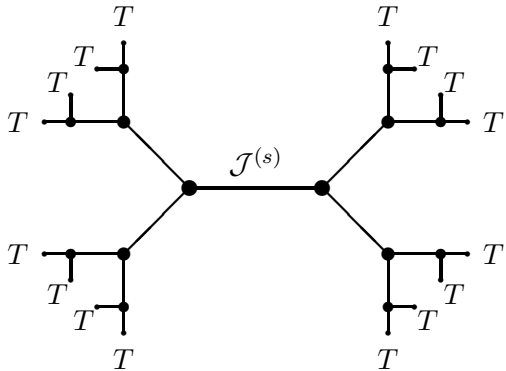
\begin{figure}[tbp]
\begin{center}
\begin{picture}(220,150)(0,10)
\thicklines

\put(50,50){\circle*{5}}
\put(50,100){\circle*{5}}
\put(75,75){\circle*{6}}
\put(125,75){\circle*{6}}
\put(150,50){\circle*{5}}
\put(150,100){\circle*{5}}

\put(50,30){\circle*{4}}
\put(30,50){\circle*{4}}

\put(50,20){\circle*{2}}
\put(40,30){\circle*{2}}
\put(30,40){\circle*{2}}
\put(20,50){\circle*{2}}

\put(30,100){\circle*{4}}
\put(50,120){\circle*{4}}

\put(20,100){\circle*{2}}
\put(30,110){\circle*{2}}
\put(40,120){\circle*{2}}
\put(50,130){\circle*{2}}

\put(170,100){\circle*{4}}
\put(150,120){\circle*{4}}

\put(180,100){\circle*{2}}
\put(170,110){\circle*{2}}
\put(160,120){\circle*{2}}
\put(150,130){\circle*{2}}

\put(150,30){\circle*{4}}
\put(170,50){\circle*{4}}

\put(160,30){\circle*{2}}
\put(150,20){\circle*{2}}
\put(180,50){\circle*{2}}
\put(170,40){\circle*{2}}

\put(75,75){\line(1,0){50}}
\put(50,20){\line(0,1){30}}
\put(50,100){\line(0,1){30}}
\put(150,20){\line(0,1){30}}
\put(150,100){\line(0,1){30}}
\put(20,50){\line(1,0){30}}
\put(150,50){\line(1,0){30}}
\put(20,100){\line(1,0){30}}
\put(150,100){\line(1,0){30}}

\put(40,30){\line(1,0){10}}
\put(40,120){\line(1,0){10}}
\put(150,30){\line(1,0){10}}
\put(150,120){\line(1,0){10}}

\put(30,100){\line(0,1){10}}
\put(30,40){\line(0,1){10}}
\put(170,100){\line(0,1){10}}
\put(170,40){\line(0,1){10}}

\put(100,85){\makebox(0,0){${\cal J}^{(s)}$}}

\put(50,10){\makebox(0,0){$T$}}
\put(25,35){\makebox(0,0){$T$}}
\put(35,25){\makebox(0,0){$T$}}
\put(10,50){\makebox(0,0){$T$}}

\put(150,10){\makebox(0,0){$T$}}
\put(175,35){\makebox(0,0){$T$}}
\put(165,25){\makebox(0,0){$T$}}
\put(190,50){\makebox(0,0){$T$}}

\put(50,140){\makebox(0,0){$T$}}
\put(25,115){\makebox(0,0){$T$}}
\put(35,125){\makebox(0,0){$T$}}
\put(10,100){\makebox(0,0){$T$}}

\put(150,140){\makebox(0,0){$T$}}
\put(175,115){\makebox(0,0){$T$}}
\put(165,125){\makebox(0,0){$T$}}
\put(190,100){\makebox(0,0){$T$}}

\put(125,75){\line(1,1){25}}
\put(125,75){\line(1,-1){25}}
\put(75,75){\line(-1,1){25}}
\put(75,75){\line(-1,-1){25}}

\end{picture}
\end{center}
\caption{Higher spin central charges from many-point functions of the stress
tensor.}
\end{figure}

Only the quantities $c_{s}^{I}$ interpolate between nonvanishing critical
values. The other quantities $c_{s,s^{\prime }}^{I},$ $s^{\prime }\leq s-2,$
typically vanish like $\beta ^{2}$ at criticality. Examples are the function 
$c_{2,0}^{T}$ or, in two dimensions, the derivative of Zamolodchikov's $c$%
-function \cite{zamolo}. These functions might be relevant to prove the $a$%
-theorems of sect. 3 and \cite{noi}.

By considering many-point functions of the stress tensor it is possible to
generate channels of arbitrary spin. Taking the limits which pairs of $T$'s
become close, one generates, for example, spin-4 operators ${\cal J}^{(4)}$.
Then taking the limit in which two ${\cal J}^{(4)}$'s become close one can
generate spin-8 operators ${\cal J}^{(8)}$ and so on. Two ${\cal J}^{(4)}$'s
also generate lower spin operators, for example some ${\cal J}^{(2)}$'s
(recall that two stress tensors produce also a spin-0 operator $\Sigma $),
which are clearly linear combinations of the known ones. In conclusion, we have
in general three central charges for each even spin greater than zero 
(vector, spinor and scalar), two central charges
for each odd spin grater than one
(vector and spinor), one central charge
for spin 0 (scalar) and one central charge for spin 1 (spinor).

\section{Outlook.}

In this paper we have worked out
a general set-up for the application of techniques similar
to those familiar in the context of deep inelastic scattering
to conformal field theories in four dimensions and quantum field theories
interpolating between pairs of them. 

The various issues that we have considered constitute in some sense the
ground level of our project, which is the investigation of quantum field
theory as a radiative interpolation between pairs of conformal field
theories. We have revealed the nature of the ``secondary central charges''
introduced in \cite{noialtri,c',ccfis}, but several questions remain
unanswered and a large part of the work remains to be done. In particular:
two-loop calculations like the one of \cite{c'} would be desirable; several
parts of the analysis presented here in four dimensions should be performed
also in the $\varepsilon \rightarrow 0$ limit starting from $4-\varepsilon $
dimensions; higher spin current OPE's and many-$T$-point functions should be
investigated; higher spin anomalies should be understood better, in
particular the quantities $a_{s}$; theories with higher spin flavor
symmetries (admitting that they exist) should be classified; 
etc. Non-perturbative arguments like e-m duality or the AdS/CFT
correspondence might prove useful to reach a better understanding of the
phenomena that we have discussed, in particular the splitting of the $I$%
-degeneracy, and maybe to compute exact IR\ values of the secondary central
charges. It would be great to achieve for the higher spin central charges
the same success achieved in \cite{noi} for the spin-1 and spin-2 ones.
Numerical computations are under consideration. Last but not least, we
mention again the main objective, which is to provide a suitable framework
for the formulation of an action principle for the RG trajectory connecting
the UV\ and the IR\ conformal limits. We will report soon on some of these
developments.

\vskip .5truecm

Acknowledgments. We would like to thank S. Ferrara, A.A. Johansen and L.
Palacios for useful correspondence and discussions. This work is partially
supported by EEC grants CHRX-CT93-0340 and TMR-516055.

\section{Appendix: higher spin tensor currents and supersymmetry.}

We use the conventions of the book {\it Superspace} \cite{superspace}.
Spinor derivatives are defined by 
\[
D_{\alpha }=\frac{\partial }{\partial \theta ^{\alpha }}+\frac{i}{2}\theta ^{%
\dot{\alpha}}\partial _{\alpha \dot{\alpha}},\qquad D_{\dot{\alpha}}=\frac{%
\partial }{\partial \theta ^{\dot{\alpha}}}+\frac{i}{2}\theta ^{\alpha
}\partial _{\alpha \dot{\alpha}}. 
\]
Chiral fields $\Phi $ satisfy $D_{\dot{\alpha}}\Phi =D_{\alpha }\bar{\Phi}=0$%
. The field equations read $\bar{D}^{2}\bar{\Phi}=D^{2}\Phi =0.$ For vector
chiral superfields we have $D_{\dot{\alpha}}W_{\alpha }=D_{\alpha }W_{\dot{%
\alpha}}=0.$ The reality condition, combined with the field equations,
imposes $D_{\dot{\alpha}}W^{\dot{\alpha}}=D_{\alpha }W^{\alpha }=0$. We
recall that $\{D_{\alpha },D_{\beta }\}=\{D_{\dot{\alpha}},D_{\dot{\beta}%
}\}=0$ and $\{D_{\alpha },D_{\dot{\alpha}}\}=i\partial _{\alpha \dot{\alpha}%
} $.

The stress tensors superfields for vector and matter multiplets can be
written as 
\[
J_{\alpha \dot{\alpha}}=W_{\alpha }W_{\dot{\alpha}},\qquad J_{\alpha \dot{%
\alpha}}=\frac{1}{3}\bar{\Phi}\overleftrightarrow{D}_{\dot{\alpha}}%
\overleftrightarrow{D}_{\alpha }\Phi . 
\]
The latter is a condensed notation that we use in this appendix. It has to
be meant as follows. One writes down all possible terms, starting from $\bar{%
\Phi}D_{\dot{\alpha}}D_{\alpha }D_{\dot{\beta}}D_{\beta }\cdots \Phi $ and
moving derivatives according only to their statistics (i.e. neglecting
(anti-)commutators) plus an additional minus sign any time a derivative is
moved from $\Phi $ to $\bar{\Phi}.$ In this process, terms like $\bar{\Phi}%
D_{\dot{\alpha}}D_{\alpha }D_{\beta }D_{\dot{\beta}}\Phi $ are not meant to
vanish, rather derivatives have to be anticommuted till they give the only
nonvanishing combination. Concretely, $\bar{\Phi}D_{\dot{\alpha}}D_{\alpha
}D_{\beta }D_{\dot{\beta}}\Phi \rightarrow -\bar{\Phi}D_{\dot{\alpha}%
}D_{\alpha }D_{\dot{\beta}}D_{\beta }\Phi .$

The spin-3 N=1 superfields are, for matter and vector multiplets
respectively, 
\begin{eqnarray*}
J_{\alpha \beta \dot{\alpha}\dot{\beta}} &=&\bar{\Phi}\overleftrightarrow{D}%
_{\dot{\alpha}}\overleftrightarrow{D}_{\alpha }\overleftrightarrow{D}_{\dot{%
\beta}}\overleftrightarrow{D}_{\beta }\Phi =\bar{\Phi}D_{\dot{\alpha}%
}D_{\alpha }D_{\dot{\beta}}D_{\beta }\Phi -2D_{\dot{\alpha}}\bar{\Phi}%
~D_{\alpha }D_{\dot{\beta}}D_{\beta }\Phi \\
&&-4D_{\alpha }D_{\dot{\alpha}}\bar{\Phi}~D_{\dot{\beta}}D_{\beta }\Phi +2D_{%
\dot{\alpha}}D_{\alpha }D_{\dot{\beta}}\bar{\Phi}~D_{\beta }\Phi +D_{\alpha
}D_{\dot{\alpha}}D_{\beta }D_{\dot{\beta}}\bar{\Phi}~\Phi , \\
J_{\alpha \beta \dot{\alpha}\dot{\beta}} &=&D_{\dot{\beta}}D_{\beta
}W_{\alpha }~W_{\dot{\alpha}}-\frac{1}{2}D_{\beta }W_{\alpha }~D_{\dot{\beta}%
}W_{\dot{\alpha}}-W_{\alpha }~D_{\beta }D_{\dot{\beta}}W_{\dot{\alpha}}.
\end{eqnarray*}

The spin-4 superfield currents are 
\begin{eqnarray*}
J_{\alpha \beta \gamma \dot{\alpha}\dot{\beta}\dot{\gamma}} &=&\bar{\Phi}%
\overleftrightarrow{D}_{\dot{\alpha}}\overleftrightarrow{D}_{\alpha }%
\overleftrightarrow{D}_{\dot{\beta}}\overleftrightarrow{D}_{\beta }%
\overleftrightarrow{D}_{\dot{\gamma}}\overleftrightarrow{D}_{\gamma }\Phi .
\\
J_{\alpha \beta \gamma \dot{\alpha}\dot{\beta}\dot{\gamma}} &=&D_{\dot{\beta}%
}D_{\beta }D_{\dot{\gamma}}D_{\gamma }W_{\alpha }~W_{\dot{\alpha}}-D_{\beta
}D_{\dot{\gamma}}D_{\gamma }W_{\alpha }~D_{\dot{\beta}}W_{\dot{\alpha}}-3D_{%
\dot{\gamma}}D_{\gamma }W_{\alpha }~D_{\beta }D_{\dot{\beta}}W_{\dot{\alpha}}
\\
&&+D_{\gamma }W_{\alpha }~D_{\dot{\gamma}}D_{\beta }D_{\dot{\beta}}W_{\dot{%
\alpha}}+W_{\alpha }~D_{\beta }D_{\dot{\beta}}D_{\gamma }D_{\dot{\gamma}}W_{%
\dot{\alpha}}
\end{eqnarray*}
Complete symmetrization in dotted and undotted indices are understood.
Decomposing these superfields one obtains automatically the improved
versions of the currents. Similar formulas can be written down for generic
spin $s$. Conservation and simple tracelessness are imposed by symmetry in
the indices and by the superfield condition $D^{\dot\alpha_i}J_{\alpha_1\dot%
\alpha_1\cdots\alpha_s\dot\alpha_s}=0$ $\forall i$.

\end{document}